\documentclass[12pt]{article}

\usepackage{amsmath}
\usepackage{amssymb}
\usepackage{physics}
\usepackage{bm}
\usepackage{comment}
\usepackage{appendix}
\usepackage{color}
\usepackage[normalem]{ulem}

\allowdisplaybreaks

\DeclareFontFamily{OT1}{pzc}{}
\DeclareFontShape{OT1}{pzc}{m}{it}{<-> s * [1.10] pzcmi7t}{}
\DeclareMathAlphabet{\mathpzc}{OT1}{pzc}{m}{it}

\usepackage{hyperref}

\usepackage{tikz,tikz-cd}
\usepackage{here}

\makeatletter
	
	\@addtoreset{equation}{section}

	\@addtoreset{figure}{section}

	\@addtoreset{table}{section}
\makeatother

\begin{document}

\baselineskip=17pt

\begin{titlepage}
\rightline{\tt KUNS-3099}

\begin{center}
\vskip 2.5cm
{\Large \bf {A Non-local ``Boundary'' Term for Two-Point Amplitudes in String Field Theory}}

\vskip 1.0cm
{\large Yasu Gen$^{a,1}$, Hiroaki Matsunaga$^{b,2}$, Jojiro Totsuka$^{a,3}$}
\vskip 1.0cm

$ ^a${\it {Department of Physics, Kyoto University,}}\\
{\it {Kitashirakawa, Kyoto 606-8502, Japan}}\\

$ ^b${\it {Osaka Metropolitan University College of Technology, \\ Saiwai-cho, Neyagawa, Osaka 572-8572, Japan}}\\

$ ^1$yasugen@gauge.scphys.kyoto-u.ac.jp\\
$ ^2$via.string.field@gmail.com\\
$ ^3$georgeyo1227@gmail.com,\\
\vspace{2mm}

\vskip 1.0cm

{\bf Abstract}
\end{center}
\emergencystretch=1em
\noindent
We propose a new “boundary” term in free bosonic string field theory. With the original choice of energy operators, the term is non-local in time. We construct this term as a bilinear form involving the commutator of the BRST operator with the step-function operator. The resulting action restores cyclicity, as required for a variational principle. When evaluated on on-shell string fields, the proposed term reproduces the correct tree-level two-point amplitude for both open and closed strings. We verify this explicitly in the tachyon and massless sectors. An alternative off-shell choice yields a time-local boundary term with the same on-shell amplitudes. For the open-string tachyon, this yields an initial-data prescription.

\end{titlepage}

\tableofcontents

\newpage
\section{Introduction and summary}

String field theory, a non-perturbative formulation of string theory, has long been developed without explicitly accounting for contributions from spacetime boundaries.
In field theory, however, boundary effects are often physically important.
This is especially true for closed string field theory, whose spectrum includes gravity, where boundary contributions can be essential.

In recent years, several works have constructed boundary terms for free string field theory \cite{Firat:2024kxq, Stettinger:2024uus, Maccaferri:2025orz, Maccaferri:2025onc}.
The basic idea can be summarized as follows.
In the absence of boundary effects, the free string field theory action is written in terms of a symplectic form defined from the BPZ inner product and a kinetic operator given by the BRST operator $Q$.
One then requires a compatibility condition between the symplectic form and $Q$, often referred to as cyclicity.
This property ensures that varying the action yields the correct equations of motion.
Once spacetime boundary contributions are taken into account, however, cyclicity is generically violated by a quantity localized on the boundary.
In order to respect the variational principle and restore the cyclicity of the full theory, one must add additional terms that cancel this localized violation.
These terms are precisely the boundary terms in the action, and \cite{Firat:2024kxq, Stettinger:2024uus, Maccaferri:2025orz, Maccaferri:2025onc} explicitly construct such boundary contributions. In particular, their constructions are compatible with Dirichlet boundary conditions.

Independent of these developments, a general framework for describing field theories on manifolds with boundary in terms of relative $L_\infty$ algebras has been proposed in \cite{Alfonsi:2024utl}.
This approach can be viewed as a homotopy-algebraic realization of the BV-BFV formalism \cite{Cattaneo:2012qu}, namely an extension of the BV formalism that incorporates boundary contributions.
In the absence of boundaries, within the homotopy algebraic framework, scattering amplitudes with three or more external states are described by the minimal model, and the two-point amplitude always vanishes.
In the framework of the relative $L_\infty$ algebra, boundary contributions are incorporated to construct a \textit{generalized} minimal model, which yields a non-vanishing two-point amplitude.
Focusing in particular on the free part, \cite{Alfonsi:2024utl} makes the following claims.
First, a field theory on a manifold with boundary is described by a degree-odd symplectic form and a kinetic operator, together with an additional degree-even bilinear form that compensates for the failure of cyclicity.
This degree-even bilinear form plays the role of the boundary term in the action.
In this respect, the structure closely parallels the mechanism emphasized in \cite{Firat:2024kxq, Stettinger:2024uus, Maccaferri:2025orz, Maccaferri:2025onc}.
Moreover, \cite{Alfonsi:2024utl} argues that for an appropriately constructed degree-even bilinear form, restricting the input fields to on-shell configurations yields the tree-level two-point amplitude.

In string theory, a tree-level $n$-point amplitude is computed as a worldsheet correlation function with $n$ insertions of on-shell vertex operators $\langle \mathcal{V}_1\cdots\mathcal{V}_n\rangle$,
on the disk for open strings and on the sphere for closed strings.
One usually fixes the insertion points of three vertex operators in order to fix the worldsheet residual symmetry.
For the two-point amplitude, this procedure does not completely fix the gauge redundancy, and the remaining unfixed symmetry has an infinite volume.
For this reason, it was traditionally argued that the tree-level two-point amplitude vanishes after dividing by this infinite volume.
It is now understood, however, that the would-be zero is compensated by divergences originating from the delta function implementing energy conservation, yielding a non-trivial finite result \cite{Erbin:2019uiz}.
The computation in \cite{Erbin:2019uiz} was carried out in the path-integral formalism, but operator-based methods have also been developed in \cite{Seki:2019ycz, Seki:2021ivm, Kishimoto:2024aig}.
In this approach, one inserts an additional operator, called the mostly BRST-exact operator 
\begin{align}
    \mathpzc{E}(z,\bar{z}):=\int_{-\infty}^{\infty} \frac{dq}{2\pi}
    \Big\{ :(c\,\partial X^0 + \tilde{c}\,\bar{\partial} X^0)\, e^{-iqX^0}(z,\bar{z}):
    - i\alpha' \frac{q}{4} (\partial c + \bar{\partial} \tilde{c})\, :e^{-iqX^0}(z,\bar{z}): \Big\}\,,\label{mostly BRST-exact operator}
\end{align}
and computes the tree-level two-point amplitude effectively as a three-point correlation function $\langle \mathpzc{E} \,\mathcal{V}_1\mathcal{V}_2\rangle$.
The quantity obtained in this way differs from the two-point amplitude of \cite{Erbin:2019uiz} by an extra factor given by the sign of the energy of the incoming string.
Thus, the overall sign is somewhat subtle in the operator formalism.
Apart from this sign factor, however, the result is consistent with \cite{Erbin:2019uiz}.

These observations suggest that one should be able to construct boundary terms in free string field theory that reproduce the tree-level two-point amplitude.
Pursuing this possibility naturally leads to boundary terms that differ from those obtained solely by imposing compatibility with Dirichlet boundary conditions, as in \cite{Firat:2024kxq, Stettinger:2024uus, Maccaferri:2025orz, Maccaferri:2025onc}.

In this paper, we propose the following action for free string field theory with a ``boundary'' term 
\begin{align}
    S=\frac{1}{2}\omega\big(\Phi,\Theta_0[X^0] Q\Phi\big)+\frac{1}{4}\omega\Big(\Phi,\,\big\{P_+,\,\big[Q,\,\Theta_0[X^0]\big]\big\}\Phi\Big)\,,\label{new action}
\end{align}
where precise definitions of the operators will be given in Section \ref{sec:preliminaries}.
Here, $\Theta_0[X^0]$ is an operator that inserts a step function of the zero-th component of the worldsheet scalar field $X^0$.
In the zero string-length limit, namely $\alpha'\to 0$, it reduces to the step function of the target-space time coordinate $x^0$, which corresponds to introducing a boundary at $x^0=0$.
The insertion of $\Theta_0[X^0]$ in the first term breaks cyclicity, and the second term is designed precisely to compensate for this violation.
The second term is controlled by the commutator $\big[Q,\Theta_0[X^0]\big]$.
This commutator, in fact, corresponds to the insertion of the mostly BRST-exact operator appearing in the operator formalism.
On shell, $P_+$ selects positive-energy modes and removes the extra sign factor that arises in the operator-formalism computation.
When $\Phi$ is restricted to on-shell states, this term reproduces the tree-level two-point amplitude up to an overall normalization.
Since $\big[Q,\Theta_0[X^0]\big]$ is non-zero only where the step function varies, it is localized near the boundary, within a region whose thickness is of order the string length.
For this reason, we refer to the second term as a ``boundary'' term. 
For the choice \eqref{P_pm}, however, $P_+$ is non-local in time, so the full additional term is not localized near the boundary.
This is why we place the phrase ``boundary'' term in quotation marks.
Section~\ref{sec:time-local} introduces an alternative off-shell choice of $P_\pm$ that is local in time and preserves the on-shell two-point amplitudes.
For the open-string tachyon, the additional term is then supported at $x^0=0$, although it remains non-local in space.
We show that imposing the equations of motion for the field and its variation only at this surface, together with a condition fixing one component of the initial data, eliminates the boundary variation.
These conditions therefore admit an initial-data interpretation.
The first term in \eqref{new action} reduces to the standard bulk free action in the limit $\alpha'\to 0$.
Interestingly, at finite $\alpha'$, additional contributions appear, and they can, in fact, be rewritten as boundary terms.
As expected, these contributions are proportional to the equations of motion and therefore do not affect on-shell physics.
At present, we do not have a definitive interpretation of these terms.
Since they are intrinsically stringy and off-shell, they may represent effects specific to string field theory.

This paper is organized as follows.
In Section \ref{sec:preliminaries}, we review free string field theory without boundaries and fix our notation and conventions. In particular, we introduce the symplectic form, the BRST operator, and the operators $\Theta_0[X^0]$ and $P_\pm$, which play a central role in our construction.
In Section \ref{sec:A new proposal for “boundary” terms in SFT}, we propose a new action for free string field theory that incorporates additional terms restoring cyclicity in the presence of the operator $\Theta_0[X^0]$. We show that the resulting action restores cyclicity and interpret the additional contribution as a ``boundary'' term.
In Section \ref{sec:example}, we analyze explicit examples in the tachyon and massless sectors for both open and closed strings. We compute the corresponding actions and verify that the ``boundary'' term correctly reproduces the tree-level two-point amplitudes.
In Section \ref{sec:time-local}, we introduce a time-local choice of $P_\pm$ and derive an initial-data prescription for the open-string tachyon.
Finally, in Section \ref{sec:Discussion}, we discuss possible future directions.

Throughout this paper, we focus on bosonic string field theory in $d=26$ dimensions. We adopt the mostly-plus convention for the target-space metric, $\eta_{\mu\nu} = \mathrm{diag}(-,+,\dots,+)$.

\section{Preliminaries}\label{sec:preliminaries}
\subsection{Free string field theory without boundaries}

We begin by reviewing the compatibility of free string field theory without spacetime boundaries with the variational principle, mainly to fix our notation and conventions.
The free action is given by
\begin{align}
S=\frac{1}{2}\omega(\Phi,\,Q\Phi)\,.\label{action without boundary}
\end{align}
The string field $\Phi$ is taken to be a general element of the first-quantized Fock space, namely a linear combination of Fock states.
For the open string, we write
\begin{align}
\Phi=\int\frac{\dd^d k}{(2\pi)^d}\, \phi^{\mu_1,\dots,\mu_r}(k)\, \alpha^{\mu_1}_{-n_1}\cdots\alpha^{\mu_r}_{-n_r}b_{-m_1}\cdots b_{-m_s}c_{-p_1}\cdots c_{-p_t}\ket{0,k}\,,\label{Phi}
\end{align}
for $n_i\geq 1\,,\quad m_i\geq 2\,,\quad p_i\geq -1$.
For the closed string, we consider Fock states that also include the anti-holomorphic modes $\tilde{\alpha}^{\tilde{\mu}}_{-\tilde{n}}$, $\tilde{b}_{-\tilde{m}}$, and $\tilde{c}_{-\tilde{p}}$ in addition to the holomorphic ones, and we impose both level matching and $(b_0-\tilde b_0)\Phi=0$.
The coefficients $\phi(k)$ are component fields, which correspond to target-space fields.
Here, a momentum dependent vacuum is defined  as
\begin{align}
    \ket{0,k}:=\,:e^{ik \cdot X}(0,0):\ket{0}=e^{ik\cdot x}\ket{0}\,,
\end{align}
where $:\ \ :$ denotes normal ordering, and the dot denotes the contraction $k\cdot X = k_\mu X^\mu$.
The mode expansion of $X^\mu(z,\bar{z})$ is given by
\begin{align}
    X^\mu(z,\bar{z})=x^\mu-i\alpha'p^\mu \log|z|^2+i\left(\frac{\alpha'}{2}\right)^{\frac{1}{2}}\sum_{\substack{m=-\infty\\m\neq0}}^{\infty}\frac{\alpha^\mu_m}{m}\left(\frac{1}{z^m}+\frac{1}{\bar{z}^m}\right)\,,
\end{align}
for openstring and
\begin{align}
    X^\mu(z,\bar{z})=x^\mu-i\frac{\alpha'}{2}p^\mu \log|z|^2+i\left(\frac{\alpha'}{2}\right)^{\frac{1}{2}}\sum_{\substack{m=-\infty\\m\neq0}}^{\infty}\frac{1}{m}\left(\frac{\alpha^\mu_m}{z^m}+\frac{\tilde{\alpha}^\mu_m}{\bar{z}^m}\right)\,,
\end{align}
for a closed string.
We define the degree of the string field $\Phi$ by ghost number minus one for the open string, and by ghost number minus two for the closed string.

The BRST operator $Q$ is defined as follows. For the open string,
\begin{align}
    Q=\sum^\infty_{n=-\infty}  
    c_{-n}L^{(\text{m})}_n
    +
    \!\!\sum^\infty_{\substack{m,n=-\infty\\m\neq n}}
    \!\!\!\frac{(m-n)}{2}
    :\!c_m c_nb_{-m-n}\!:
    \,-
    \,c_0\,,\label{BRST open}
\end{align}
and for the closed string,
\begin{align}
    Q=\sum^\infty_{n=-\infty}  
    \left(c_{-n}L^{(\text{m})}_n+\tilde{c}_{-n}\tilde{L}_n^{(\text{m})}\right)
    +
    \!\!\sum^\infty_{\substack{m,n=-\infty\\m\neq n}}
    \!\!\!\frac{(m-n)}{2}
    :c_m c_nb_{-m-n}\,+\,\tilde{c}_m \tilde{c}_n\tilde{b}_{-m-n}\!:
    \,-
    \,(c_0+\tilde{c}_0)\,.\label{BRST closed}
\end{align}
Here $L_n^{(m)}$ and $\tilde{L}_n^{(m)}$ denote the Virasoro operator in the matter sector.

The symplectic form $\omega$ is defined from BPZ inner product $\langle\cdot,\cdot\rangle$ by
\begin{align}
    \omega(\Phi_1,\,\Phi_2):=(-1)^{\Phi_1} \langle \Phi_1, \#\,\Phi_2 \rangle\,,\label{def of omega}
\end{align}
where $\#$ represents a ghost operator that is inserted when required:
\begin{align}
    \#=\begin{cases}
        \ c_0^-=\frac{1}{2}(c_0-\tilde{c}_0)&\,,\quad \text{for closed strings}\,,\\
        \ 1 &\,,\quad \text{for open strings}\,.\label{c insertion}
    \end{cases}
\end{align}
The exponent of the sign factor on the right-hand side of \eqref{def of omega} is the degree of $\Phi_1$.
This symplectic form is graded anti-symmetric:
\begin{align}
    \omega(\Phi_1,\,\Phi_2)=-(-1)^{\Phi_1\Phi_2}\omega(\Phi_2,\,\Phi_1)\,.
\end{align}
We normalize the BPZ inner product as follows. 
For the open string,
\begin{align}
\bra{0,k_1}c_{-1}c_0c_1\ket{0,k_2}=(2\pi)^d\delta^{(d)}(k_1+k_2)\,,
\end{align}
while for the closed string,
\begin{align}
\bra{0,k_1}c_{-1}c_0c_1\tilde{c}_{-1}\tilde{c}_0\tilde{c}_1\ket{0,k_2}=(2\pi)^d\delta^{(d)}(k_1+k_2)\,,
\end{align}
where $\bra{0,k}$ denotes the BPZ conjugate of $\ket{0,k}$.

The symplectic form $\omega$ and the BRST operator $Q$, defined as above, satisfy the property known as cyclicity:
\begin{align}
    \omega(\Phi_1,\,Q\Phi_2)=(-1)^{\Phi_1+\Phi_2+\Phi_1\Phi_2}\,\omega(\Phi_2,\,Q\Phi_1)\,.\label{cyclicity of Q}
\end{align}
From this property, it follows that the string field $\Phi$ appearing in the action \eqref{action without boundary} must have degree even; otherwise, the action vanishes.
Moreover, this property allows us to compute the variation of the action as
\begin{align}
\delta S=\omega(\delta\Phi,\,Q\Phi)
\end{align}
and hence yields the equation of motion $Q\Phi=0$.
In other words, cyclicity is the condition required for the variational principle to be well defined.

\subsection{Definitions of $\Theta_0[X^0]$ and $P_\pm$}\label{Sec:Definitions of some operators}

In this subsection, as preparation for introducing the main result of this paper, namely the action \eqref{new action}, we define the operators $\Theta_0[X^0]$ and $P_+$.

First, we define $\Theta[X^0](z,\bar z)$ by
\begin{align}
\Theta[X^0](z,\bar{z}):=\int_{-\infty}^\infty \frac{\dd q}{2\pi} \frac{i}{q}:e^{-iqX^0}(z,\bar{z}):+\frac{1}{2}\, .\label{Theta(X)}
\end{align}
In the limit $\alpha'\to 0$, $X^0$ reduces to the target-space time coordinate $x^0$, and in this limit the operator becomes the step function,
\begin{align}
    \theta(x^0):=\int_{-\infty}^\infty \frac{\dd q}{2\pi} \frac{i}{q}e^{-iqx^0}+\frac{1}{2}=
    \begin{cases}
        1\,,\quad x^0>0\,,\\
        \frac{1}{2}\,,\quad x^0=0\\
        0\,,\quad x^0<0\,.
    \end{cases}
\end{align}
Thus, $\Theta[X^0](z,\bar z)$ is a stringy extension of this step function.

We define $\Theta_0[X^0]$ as the contour integral of this operator along the unit circle $\mathcal{C}$ with $|z|=1$.
For the open string,
\begin{align}
\Theta_0[X^0]:=\oint_{\mathcal{C}}\frac{\dd z}{2\pi i}\frac{1}{z}\,\Theta[X^0](z,\bar{z})\,,
\end{align}
and for the closed string,
\begin{align}
\Theta_0[X^0]:=\frac{1}{2}\Big(\oint_{\mathcal{C}}\frac{\dd z}{2\pi i}\frac{1}{z}-\oint_{\mathcal{C}}\frac{\dd \bar{z}}{2\pi i}\frac{1}{\bar{z}}\Big)\,\Theta[X^0](z,\bar{z})\,.
\end{align}
Here we use the convention $\oint_\mathcal{C}\frac{\dd z}{2\pi i}\frac{1}{z}=-\oint_\mathcal{C}\frac{\dd \bar z}{2\pi i}\frac{1}{\bar z}=1$.
Since this operator is integrated along the unit circle $|z|=1$, it is invariant under BPZ conjugation. Namely, it satisfies
\begin{align}
\omega\big(\Phi_1,\,\Theta_0[X^0]\Phi_2\big)=\omega\big(\Theta_0[X^0]\Phi_1,\,\Phi_2\big)\,.\label{cyclicity Theta}
\end{align}

Computing the commutator of $\Theta[X^0](z,\bar z)$ with the BRST operator, we find
\begin{align}
\big[Q,\Theta[X^0](z,\bar{z})\big]=\int_{-\infty}^{\infty} \frac{dq}{2\pi}
    \Big\{ :(c\,\partial X^0 + \tilde{c}\,\bar{\partial} X^0)\, e^{-iqX^0}(z,\bar{z}):
    - i\alpha' \frac{q}{4} (\partial c + \bar{\partial} \tilde{c})\, :e^{-iqX^0}(z,\bar{z}): \Big\}
\end{align}
which coincides with the mostly BRST-exact operator $\mathpzc{E}(z,\bar{z})$ \eqref{mostly BRST-exact operator} in \cite{Seki:2019ycz, Seki:2021ivm, Kishimoto:2024aig}\footnote{In \cite{Seki:2019ycz, Seki:2021ivm}, the mostly BRST-exact operator appears as a boundary operator, while in \cite{Kishimoto:2024aig} it appears as a bulk operator. The operator used in this paper is of the latter type.}.
Therefore, the commutator $\big[Q,\Theta_0[X^0]\big]$ appearing in the action \eqref{new action} is the contour integral of this mostly BRST-exact operator.

We make an important comment.
The integrand defining $\Theta[X^0](z,\bar z)$ diverges at $q=0$, and hence it is not a well-defined operator on the full state space.
By contrast, the mostly BRST-exact operator $\mathpzc{E}(z,\bar z)$ does not contain such a divergence and is well defined.
Since $\mathpzc{E}(z,\bar z)=\big[Q,\Theta[X^0](z,\bar z)\big]$, this operator is BRST-exact for $q\neq 0$.
For $q=0$, however, $\Theta[X^0](z,\bar z)$ is ill defined, and $\mathpzc{E}(z,\bar z)$ is therefore not BRST-exact.
This is the reason it is called mostly BRST-exact.

Next, we define the operator $P_+$.
Let $\Phi(k)$ denote the part of the expansion \eqref{Phi} labeled by the momentum $k$.
Namely,
\begin{align}
\Phi=\int\frac{\dd^d k}{(2\pi)^d}\,\Phi(k)\,.
\end{align}
We define $P_\pm$ by
\begin{align}
P_\pm \Phi=\int\frac{\dd^d k}{(2\pi)^d}\,\theta(\pm k^0)\,\Phi(k)\,,\label{P_pm}
\end{align}
where $\theta(k^0)$ is the step function of the zero-th component of the momentum, namely the energy, defined by
\begin{align}
    \theta(k^0):=
    \begin{cases}
        1\quad k^0>0\,,\\
        \frac{1}{2}\quad k^0=0\,,\\
        0\quad k^0 < 0\,.
    \end{cases}
\end{align}
That is, $P_+$ selects positive-energy modes, while $P_-$ selects negative-energy modes.
The definition \eqref{P_pm} is used through Section~\ref{sec:boundary condition}; an alternative off-shell choice is introduced in Section~\ref{sec:time-local}.
These operators satisfy
\begin{align}
    P_++P_-&=\mathbb{I}\,,\label{P+P-}\\
    P_+-P_-&=\mathrm{sgn}(p^0)\,,\label{sgn}
\end{align}
where $\mathbb{I}$ is the identity operator and $\mathrm{sgn}(p^0)$
denotes the sign operator acting on
\begin{align}
    \mathrm{sgn}(p^0) \Phi(k):=\frac{k^0}{|k^0|}\,\Phi(k)\,.\label{def of sgn}
\end{align}
An important property is that the symplectic form $\omega$ defined by \eqref{def of omega} satisfies
\begin{align}
\omega(P_+ \Phi_1,\,\Phi_2)=\omega(\Phi_1,\,P_- \Phi_2)\,.\label{omega P_pm}
\end{align}
This corresponds to the energy conservation.

With these preparations, in the next section, we propose a new action for free string field theory including boundary contributions.

\section{A new proposal for ``boundary'' terms in string field theory}\label{sec:A new proposal for “boundary” terms in SFT}

Using the operators introduced above, we propose the following new action for free string field theory with boundary contributions:
\begin{align}
S=\frac{1}{2}\omega\big(\Phi,\Theta_0[X^0] Q\Phi\big)+\frac{1}{4}\omega\Big(\Phi,\,\big\{P_+,\,\big[Q,\,\Theta_0[X^0]\big]\big\}\Phi\Big)\,.\label{new action 2}
\end{align}
In this section, we show that cyclicity is restored for this action, and that the ``boundary'' term reproduces the correct tree-level two-point amplitude.
This two-point amplitude corresponds to the quadratic part of the generalized minimal model introduced in \cite{Alfonsi:2024utl}.

\subsection{Restoring cyclicity}\label{sec:Restoring cyclicity}

In contrast to the fact that the BRST operator $Q$ is cyclic with respect to the original symplectic form $\omega$, it is not cyclic with respect to the modified symplectic form $\omega\big(\cdot,\Theta_0[X^0]\cdot\big)$ with an insertion of $\Theta_0[X^0]$.
The violation is calculated as
\begin{align}
    \omega\big(\Phi_1,\,\Theta_0[X^0]Q\Phi_2\big)-(-1)^{\Phi_1+\Phi_2+\Phi_1\Phi_2}\omega\big(\Phi_2,\,\Theta_0[X^0]Q\Phi_1\big)=-\omega\big(\Phi_1,\,\big[Q,\,\Theta_0[X^0]\big]\Phi_2\big)\,.
\end{align}
Because of this violation of cyclicity, the first term of the action \eqref{new action 2} alone does not admit a well-defined variational principle.
As a general statement, suppose that there exists an operator $\mathcal{B}$ such that
\begin{align}
\big[Q,\Theta_0[X^0]\big]=\mathcal{B}+\mathcal{B}^*,\label{def of B}
\end{align}
where $\mathcal{B}^*$ is the adjoint of $\mathcal{B}$ with respect to $\omega$, namely it satisfies
\begin{align}
\omega(\Phi_1,\,\mathcal{B}\Phi_2)=(-1)^{\Phi_1}\omega(\mathcal{B}^*\Phi_1,\,\Phi_2).
\end{align}
Then, cyclicity can be restored in the form
\begin{align}
\omega\big(\Phi_1,\,\Theta_0[X^0]Q\Phi_2\big)+\omega(\Phi_1,\mathcal{B}\Phi_2)
= (-1)^{\Phi_1+\Phi_2+\Phi_1\Phi_2}\Big(
\omega\big(\Phi_2,\,\Theta_0[X^0]Q\Phi_1\big)+\omega(\Phi_2,\,\mathcal{B}\Phi_1)
\Big).
\end{align}
Therefore,
\begin{align}
S=\frac{1}{2}\omega\big(\Phi,\,\Theta_0[X^0]Q\Phi\big)+\frac{1}{2}\omega(\Phi,\,\mathcal{B}\Phi)\,,
\end{align}
defines an action that admits a well-defined variational principle.
Note that $\mathcal{B}$ is not unique, and it is chosen so as to be compatible with the boundary conditions.

The choice
\begin{align}
\mathcal{B}=\frac{1}{2}\big\{P_+,\,\big[Q,\Theta_0[X^0]\big]\big\}\label{proposal B}
\end{align}
is one solution to \eqref{def of B}.
Indeed, using \eqref{cyclicity of Q}, \eqref{cyclicity Theta}, and \eqref{omega P_pm}, we find
\begin{align}
\mathcal{B}^*=\frac{1}{2}\big\{P_-,\,\big[Q,\,\Theta_0[X^0]\big]\big\}\,.
\end{align}
Furthermore, using \eqref{P+P-} we obtain 
\begin{align}
    \mathcal{B}+\mathcal{B}^*=\frac{1}{2}\big\{(P_++P_-),\,\big[Q,\,\Theta_0[X^0]\big]\big\}=\big[Q,\,\Theta_0[X^0]\big]\,.
\end{align}
Note that $P_\pm$ need not be defined specifically as in \eqref{P_pm}; it is sufficient that they satisfy \eqref{P+P-} and \eqref{omega P_pm}.
In particular, $P_\pm^2=P_\pm$ is not required off shell; such an
alternative choice is studied in Section~\ref{sec:time-local}.
However, the choice $P_\pm = \tfrac{1}{2}$ is not allowed.
In this case, one finds $\mathcal{B} = \mathcal{B}^* = \tfrac{1}{2}\big[Q,\Theta_0[X^0]\big]$, for which the boundary term vanishes due to \eqref{cyclicity of Q} and \eqref{cyclicity Theta}.
We emphasize that this construction of $\mathcal{B}$ does not rely on the explicit form of $Q$ or $\Theta_0[X^0]$.
It therefore provides a general method to construct a term that restores cyclicity, and can be useful beyond the present setting.
However, it is not immediately clear from this expression which ``boundary'' conditions\footnote{For the choice \eqref{P_pm}, $\mathcal{B}$ is non-local in time, so the word \textit{boundary} is used in quotation marks. The time-local alternative is considered in Section~\ref{sec:time-local}.} this term is compatible with.
After working out explicit examples, we discuss the limitation of this choice in Section~\ref{sec:boundary condition} and examine an initial-data prescription based on a time-local choice in Section~\ref{sec:time-local}.

As already mentioned in the Introduction, we refer to the second term in the action \eqref{new action 2} as a ``boundary'' term. Strictly speaking, however, this terminology is not precise.
For the choice \eqref{P_pm}, since $\theta(k^0)$ can be written in an
integral representation that involves the time coordinate over the entire range from $-\infty$ to $+\infty$, the corresponding contribution is not localized in $x^0$.
In this respect, it differs from a boundary term in the usual sense.
We therefore use quotation marks for the choice \eqref{P_pm}, whose additional term is not localized in time.
This terminology distinguishes it from the genuine boundary term obtained for the open-string tachyon with the time-local choice in Section~\ref{sec:time-local}.

\subsection{``Boundary'' term produces a two-point amplitude}\label{sec:``Boundary term'' produces two point amplitude}

The tree-level two-point amplitude can be computed from the ``boundary'' term of the action \eqref{new action 2}.
Concretely, it can be evaluated as follows.
First, we denote by $P_{\mathrm{on}}$ the projection onto on-shell states, namely onto the BRST cohomology.
It satisfies
\begin{align}
P_{\mathrm{on}}Q=QP_{\mathrm{on}}=0.
\end{align}
The coefficient of the component fields in the ``boundary'' term restricted to on-shell states,
\begin{align}
\frac{1}{4}\omega\Big(P_\mathrm{on}\Phi,\,\big\{P_+,\,\big[Q,\,\Theta_0[X^0]\big]\big\}P_\mathrm{on}\Phi\Big)\,,
\end{align}
is proportional to the tree-level two-point amplitude of the corresponding component fields.
That is, for a component field $\phi(k)$ with momentum $k^\mu$, we denote the corresponding functional derivative by $\frac{\delta}{\delta\phi(k)}$.
Then the tree-level two-point amplitude for $\phi_1(k_1)$ and $\phi_2(k_2)$ is given by
\begin{align}
\frac{C}{4}\,\frac{\delta}{\delta\phi_1(k_1)}\frac{\delta}{\delta\phi_2(k_2)}\omega\Big(P_\mathrm{on}\Phi,\,\big\{P_+,\,\big[Q,\,\Theta_0[X^0]\big]\big\}P_\mathrm{on}\Phi\Big)\,.\label{two pt amp boundary}
\end{align}
where $C$ is an overall normalization constant, to be fixed by explicit examples in the next section.

The statement that the expression \eqref{two pt amp boundary} reproduces the tree-level two-point amplitude can be understood from the operator-formalism computation of the two-point amplitude in \cite{Seki:2019ycz, Seki:2021ivm, Kishimoto:2024aig}.
In this formalism, for on-shell vertex operators $\mathcal{V}_1(z_1,\bar z_1)$ and $\mathcal{V}_2(z_2,\bar z_2)$, a three-point interaction with an additional insertion of the mostly BRST-exact operator $\mathpzc{E}(z,\bar{z})$ \eqref{mostly BRST-exact operator}
reproduces the tree-level two-point amplitude $\mathcal{A}_2$, up to an overall sign factor:
\begin{align}
&\langle\mathpzc{E}(z,\bar{z})\,\#\mathcal{V}_1(z_1,\bar z_1)\,\mathcal{V}_2(z_2,\bar z_2)\rangle\propto \frac{k_2^0}{|k_2^0|}\times \mathcal{A}_2\,, 
\label{seki-takahashi}\\
&\mathcal{A}_2:=2|k_2^0|(2\pi)^{d-1}\delta^{(d-1)}(\vb{k}_1+\vb{k}_2)\,,
\end{align}
where $\vb{k}_i$ denotes the spatial components of $k_i^\mu$. Here $k_i^\mu$ is the momentum carried by the target-space field associated with the vertex operator $\mathcal{V}_i$, and it satisfies the on-shell condition $(k_i^0)^2-(\vb{k}_i)^2=(m_i)^2$, where $m_i$ is the corresponding mass.
For the open string, $\mathcal{V}_i$ is a ghost-number-one boundary operator and can be written as $cV_i$ in terms of a matter operator $V_i$.
For the closed string, $\mathcal{V}_i$ is a ghost-number-two bulk operator and can be written as $c\tilde{c}V_i$.
The symbol $\#$ is present only in the closed-string case, where it denotes an additional insertion of $\bar{\partial}\tilde{c}-\partial c$ at the same point as $\mathcal{V}_1$.
The mostly BRST-exact operator is inserted in the bulk, and the result does not depend on the insertion points of these three operators.
This correlator is non-vanishing only when the two vertex operators correspond to the same target-space field, including the same internal labels such as spin and Lorentz indices, and differ only in their energy and momentum. In particular, it is non-zero only when energy is conserved, $k_1^0+k_2^0=0$.
The normalization here depends on whether we consider the open or closed string, but in the present discussion, this dependence is absorbed into the overall constant $C$ in \eqref{two pt amp boundary}, and does not play a role.

To view the expression \eqref{two pt amp boundary} from this perspective, we first rewrite it as follows.
Let us set
\begin{align}
\frac{\delta}{\delta\phi_i(k_i)}\Phi=\Psi_i(k_i)\,.
\end{align}
The relation to the vertex operators is given by $\Psi_i(k_i)=\mathcal{V}_i(0,0)\ket{0}$ and the string field is expressed as $\Phi=\sum_i\int\frac{\dd^d k_i}{(2\pi)^d} \phi_i(k_i) \Psi_i(k_i)$.
Then, using \eqref{sgn} and \eqref{def of sgn}, the expression \eqref{two pt amp boundary} can be written as
\begin{align}
\frac{C}{4}\left(\frac{k_2^0}{|k_2^0|}-\frac{k_1^0}{|k_1^0|}\right)\omega\Big(\Psi_1(k_1),\,\big[Q,\,\Theta_0[X^0]\big]\Psi_2(k_2)\Big)\,.\label{two pt amp rewritten}
\end{align}
Here, the energy and spatial momentum satisfy the on-shell condition $(k_i^0)^2-(\vb{k}_i)^2=(m_i)^2$.
As explained in Section \ref{sec:preliminaries}, $\big[Q,\Theta_0[X^0]\big]$ is the contour integral of the mostly BRST-exact operator $\mathpzc{E}(z,\bar{z})$.
Therefore, $\omega(\Psi_1,\big[Q,\Theta_0[X^0]\big]\Psi_2)$ reduces to a three-point correlator with an insertion of $\mathpzc{E}$.
Moreover, since the other two vertex operators are on-shell, the correlator is independent of the insertion points of the operators.
As a result, the contour integral simply picks up the poles in $1/z$ and $1/\bar z$, and yields a factor of $1$.
Furthermore, in the closed-string case, the operator $c_0^-$ built into the definition of $\omega$ in \eqref{c insertion} corresponds to an insertion of $\bar{\partial}\tilde{c}-\partial c$.
Putting these ingredients together, one finds that
\begin{align}
\omega\Big(\Psi_1(k_1),\,\big[Q,\,\Theta_0[X^0]\big]\Psi_2(k_2)\Big)\propto\frac{k^0_2}{|k^0_2|}\times \mathcal{A}_2\,.\label{seki-takahashi reder}
\end{align}
The tree-level two-point amplitude $\mathcal{A}_2$ is non-zero only when $k_1^0+k_2^0=0$.
In this case, the energy-dependent sign factor in \eqref{two pt amp rewritten} cancels the sign factor $\frac{k_2^0}{|k_2^0|}$ on the right-hand side of \eqref{seki-takahashi reder}, and hence \eqref{two pt amp boundary} reproduces the tree-level two-point amplitude:
\begin{align}
    \frac{C}{4}\,\frac{\delta}{\delta\phi_1(k_1)}\frac{\delta}{\delta\phi_2(k_2)}\omega\Big(P_\mathrm{on}\Phi,\,\big\{P_+,\,\big[Q,\,\Theta_0[X^0]\big]\big\}P_\mathrm{on}\Phi\Big) = \mathcal{A}_2\,. \label{3.18}
\end{align}
The proportionality constant was absorbed into $C$.

The claim that the tree-level two-point amplitude can be computed from a boundary term has been discussed in \cite{Alfonsi:2024utl} within the framework of relative $L_\infty$ algebras.
Our result provides an explicit example of this claim in string field theory.
In the examples presented in \cite{Alfonsi:2024utl}, however, the boundary term is chosen to be compatible with Dirichlet boundary conditions, and the extraction of the two-point amplitude, in fact, requires introducing a regularization and manually picking up positive- or negative-energy modes.
By contrast, our choice of ``boundary'' term reproduces the two-point amplitude in a more natural manner.

\section{Examples}\label{sec:example}
In this section, the free string field theory action with ``boundary'' term proposed in this paper is analysed explicitly in the tachyon and massless sectors, for both open and closed strings.
Throughout this section, $P_\pm$ are given by \eqref{P_pm}.
A common feature of the examples below is that some of the terms which appear to be localized at $x^0=0$ contain the operator $P_+$, and are therefore non-local. 
This non-locality is important for the variational principle, since it makes the corresponding ``boundary'' conditions different from ordinary Dirichlet or Neumann boundary conditions. 
We return to this issue in Section~\ref{sec:boundary condition}. The time-local alternative and its initial-data interpretation are discussed in Section~\ref{sec:time-local}.

\subsection{Open string tachyon}\label{sec:Open string tachyon}
Including the present subsection, in what follows we explicitly derive the action and the two-point amplitude using the string field, symplectic form, and step function introduced in Section~\ref{sec:Restoring cyclicity}. 
The bulk part of the resulting action agrees with that obtained in \cite{Maccaferri:2025onc}, namely, the standard bulk term with a boundary introduced by a step function. 
The additional term produced by the second term in \eqref{new action 2} has a more subtle structure: although it contains a factor of $\delta(x^0)$, it also involves the operator $P_+$ and is therefore not a local boundary term in the usual sense. 
We first show, however, that this term nevertheless reproduces the tree-level two-point amplitude correctly.

We write the kinetic term and ``boundary'' term of the action as 
\begin{align}
S=
\frac{1}{2}\omega \!\left(
\,\Phi,\,\Theta_0[X^0] Q\,\Phi
\right)
+
\frac{1}{4}\omega\big(\
\Phi,
\left\{
P_+,\left[Q,\Theta_0[X^0]\right]
\right\}\,\Phi
\big)\,.\label{action section3}
\end{align}
Here, Q is the BRST operator introduced previously in  \eqref{BRST open}, given by
\begin{align}
    Q=\sum^\infty_{n=-\infty}  
    c_{-n}L^{(\text{m})}_n
    +
    \!\!\sum^\infty_{\substack{m,n=-\infty\\m\neq n}}
    \!\!\!\frac{(m-n)}{2}
    :\!c_m c_nb_{-m-n}\!:
    \,-
    \,c_0\,.\label{BRST open Sec4}
\end{align}
In the open string tachyon sector, the string field $\Phi$ takes the form
\begin{align}
\Phi=\int \frac{\dd^{d} k}{(2\pi)^d}\,T(k)\,c_1\ket{0,k}=\int \frac{\dd^{d} k}{(2\pi)^d}\, T(k)\,c:e^{ik \cdot X}(0,0):\ket{0,0}\,.
\end{align}
For this string field and the open-string BRST operator \eqref{BRST open Sec4}, we evaluate the action \eqref{new action 2}.
First, the action of $Q$ on $c_1\ket{0,k}$ is given by
\begin{align}
    Q\,c_1\ket{0,k}=(L_0^{(m)}-1)c_0c_1\ket{0,k}=(\alpha' k^2-1)c_0c_1\ket{0,k}\,.
\end{align}
Next, we consider the action of $\Theta_0[X^0]$ on $\ket{0,k}$:
\begin{align}
    \Theta_0[X^0]\ket{0,k}=\Big(\oint_\mathcal{C}\frac{\dd z}{2\pi i}\frac{1}{z}\int_{-\infty}^\infty \frac{\dd q}{2\pi}\frac{i}{q}:e^{-iqX^0}(z,\bar{z}):+\frac{1}{2}\Big):e^{ik_\mu X^\mu}(0,0):\ket{0,0}\,.
\end{align}
The OPE of $:e^{-iqX^0}(z,\bar z):$ with $:e^{ik \cdot X}(0,0):$ is
\begin{align}
:e^{-iqX^0}(z,\bar z)::e^{ik \cdot X}(0,0):=|z|^{4\alpha' k\cdot \hat q}\,\Big(:e^{i(k+\hat q)\cdot X}(0,0):+\mathcal{O}(z)\Big) .
\end{align}
Here $\hat q$ is the vector whose zero-th component is $q$ and whose other components are all zero. Using this relation, one can show
\begin{align}
    \Theta_0[X^0]\ket{0,k}&=\oint_\mathcal{C}\frac{\dd z}{2\pi i}\frac{1}{z}\int_{-\infty}^\infty\frac{\dd q}{2\pi}\frac{i}{q}|z|^{4\alpha' k\cdot \hat q}\ket{0,k+\hat{q}}+\frac{1}{2}\ket{0,k}\notag\\
    &=\frac{1}{2}\ket{0,k}+\int_{-\infty}^\infty\frac{\dd q}{2\pi}\frac{i}{q}\ket{0,k+\hat{q}}\,.
\end{align}
Here we have omitted states at other levels, which do not contribute to the quadratic action in the sector considered.
Therefore, the following bulk contribution in the action can be computed
\begin{align}
    \Theta_0[X^0]Qc_1\ket{0,k}&=(\alpha'k^2-1)c_0c_1\Theta_0[X^0]\ket{0,k}\notag\\
    &=(\alpha'k^2-1)c_0c_1\Big(\frac{1}{2}\ket{0,k}+\int_{-\infty}^\infty\frac{\dd q}{2\pi}\frac{i}{q}\ket{0,k+\hat{q}}\Big)\,.
\end{align}
One can also compute the part appearing in the ``boundary'' term that corresponds to the mostly BRST-exact operator:
\begin{align}
    \big[Q,\Theta_0[X^0]\big]c_1\ket{0,k}&=Q\Big(\frac{1}{2}c_1\ket{0,k}+\int_{-\infty}^\infty\frac{\dd q}{2\pi}\frac{i}{q}c_1\ket{0,k+\hat{q}}\Big)-\Theta_0[X^0]\Big((\alpha' k^2-1)c_0c_1\ket{0,k}\Big)\notag\\
    &=\frac{1}{2}(\alpha'k^2-1)c_0c_1\ket{0,k}+\int_{-\infty}^\infty\frac{\dd q}{2\pi}\frac{i}{q}\Big(\alpha'(k+\hat{q})^2-1\Big)c_0c_1\ket{0,k+\hat{q}}\notag\\
    &\hspace{4cm}-(\alpha'k^2-1)c_0\Big(\frac{1}{2}c_1\ket{0,k}+\int_{-\infty}^\infty\frac{\dd q}{2\pi}\frac{i}{q}c_1\ket{0,k+\hat{q}}\Big)\notag\\
    &=\int_{-\infty}^\infty\frac{\dd q}{2\pi}i\alpha'(-2k^0-q)c_0c_1\ket{0,k+\hat{q}}\,.
\end{align}
Here we use the fact that $\Theta_0[X^0]$ contains no worldsheet ghost fields and thus commutes with $c_n,b_n$. 

Using these results, the action can be computed as
\begin{align}
\hspace{-3.5cm}S=&\,
\frac{1}{2}\omega \!\left(\Phi,\,\Theta_0[X^0] Q\,\Phi
\right)
+
\frac{1}{4}\omega\big(\Phi,
\left\{P_+,\left[Q,\Theta_0[X^0]\right]
\right\}\,\Phi
\big)\,\notag
\end{align}
\vspace{-1.5em}
\begin{align}
=& \,\,\frac{1}{2} \!\int _{k_1} \!\int _{k_2}
      T(k_1) T(k_2) 
      \bra{0,k_1} c_{-1}c_0 c_1\left(\alpha'k_2^2-1\right)\left(\int_q\frac{i}{q}\ket{0,k_2+\hat{q}}
      +\frac{1}{2}\ket{0,k_2}\right)  \notag\\[6pt]
 &\hspace{1cm}+
   \frac{1}{2} \!\!\int _{k_1}\!\int _{k_2} 
      T(k_1) T(k_2) 
      \bra{0,k_1} c_{-1}c_0c_1 
      \int_q \frac{\theta(k_2^0+q)+\theta(k_2^0)}{2}i\alpha'\left(-2k_2^0-q\right)\ket{0,k_2+\hat{q}}\,\notag\\[6pt]
      =&\,\,\frac{1}{2}\int_{k_1}\!\int_{k_2}T(k_1)T(k_2)
      \Big[\!\!\int \!\!d^dx \,\theta(x^0)\left(\alpha'k_2^2-1\right)e^{i(k_1+k_2)\cdot x}\notag\\[3pt]
      &\hspace{6cm}+i\!
      \int d^dx \,\delta(x^0)\frac{\theta(k_2^0)+\theta(-k_1^0)}{2}\alpha'(k_1^0-k_2^0)\,e^{i(k_1+k_2)\cdot x}\,\Big]\,\,.\label{open tachyon action}
\end{align}
Here we introduced the following shorthand notation, which will be used throughout the following discussion:
\begin{align}
    \int_k:=\int\!\!\frac{d^dk}{(2\pi)^d}\,,\qquad \int_q:=\int^\infty_{-\infty}\!\frac{dq}{2\pi}\,.
\end{align}
The first term on the right-hand side of \eqref{open tachyon action} is the standard bulk term for the tachyon and has support only for $x^0>0$.
The second term, on the other hand, contains $\delta(x^0)$ and therefore appears at first sight to be localized at the boundary $x^0=0$.
However, it is in fact not a local term, as will be discussed in detail in Section~\ref{sec:boundary condition}.

For the moment, let us verify that the correct two-point amplitude is reproduced from this ``boundary'' term. According to the discussion in Section~\ref{sec:``Boundary term'' produces two point amplitude}, the two-point amplitude can be obtained by substituting the on-shell fields into the ``boundary'' term with normalization constant $C$:
\begin{align}
    \mathcal{A}_2=&\frac{C}{4}\,\frac{\delta}{\delta T(k_1)}\frac{\delta}{\delta T(k_2)}\omega\Big(P_\mathrm{on}\Phi,\,\big\{P_+,\,\big[Q,\,\Theta_0[X^0]\big]\big\}P_\mathrm{on}\Phi\Big)\nonumber\\
    =&C\times\frac{\alpha'}{2i}(k^0_2-k^0_1)\left[\frac{\theta(k_2^0)+\theta(-k_1^0)}{2}-\frac{\theta(k_1^0)+\theta(-k_2^0)}{2}\right](2\pi)^{d-1}\delta^{(d-1)}(\bm{k_1}+\bm{k_2})\,.\label{open tachyon amp 1}
\end{align}
Delta function $\delta^{(d-1)}(\bm{k}_1+\bm{k}_2)$ expresses conservation of spatial momentum.
In addition, since the external states are on shell, their energies satisfy $(k_i^0)^2-(\bm{k}_i)^2=-1/\alpha'$ for $i=1,2$.
It follows that $k_1^0-k_2^0=0$ or $k_1^0+k_2^0=0$.
However, \eqref{open tachyon amp 1} contains the factor $(k_2^0-k_1^0)$, and therefore it is non-vanishing only when the energies satisfy $k_1^0+k_2^0=0$.
In this case, \eqref{open tachyon amp 1} can be further rewritten as
\begin{align}
    \mathcal{A}_2=&C\times \frac{\alpha'}{2i}\times 2k_2^0\big[\theta(k_2^0)-\theta(-k_2^0)\big](2\pi)^{d-1}\delta^{(d-1)}(\bm{k_1}+\bm{k_2})\nonumber\\
    =&C\times \frac{\alpha'}{2i}\times 2|k_2^0|(2\pi)^{d-1}\delta^{(d-1)}(\bm{k_1}+\bm{k_2})\,.\label{open tachyon amp}
\end{align}
Here, in the second equality, we used $\theta(k_2^0)-\theta(-k_2^0)=\frac{k_2^0}{|k_2^0|}$.
It follows that, in order to reproduce the correct two-point amplitude, the normalization constant $C$ should be chosen as
\begin{align}
    C=\frac{2i}{\alpha'}\,,\label{factor C in open}
\end{align}
for open string.

\subsection{Massless open string}
We will similarly derive the action and two-point amplitude for the massless open string. Substitute the string field, including the auxiliary fields,
\begin{align}
    \Phi=\!\int \!\!\frac{d^dk}{(2\pi)^d}\Big(A_\mu(k)\alpha^\mu_{-1} c_1\ket{0,k}\,-\,i\sqrt{\frac{\alpha'}{2}}B(k)c_0\ket{0,k}\Big)
\end{align}
into the action. To keep the paper self-contained, we do not present the details of the calculation, which are given in the appendix. Using the results of \eqref{massless open bulk} and \eqref{massless open boundary}, the full action takes the following form:
\begin{align}
    S=&\frac{1}{2}\omega \!\left(\,\Phi,\,\Theta_0[X^0] Q\,\Phi\right)\,+\,\frac{1}{2}\omega(\,\Phi,\tfrac{1}{2}\left\{P_+,\big[Q,\Theta_0[X^0]\big]\right\}\,\Phi)\notag\\
    =&\frac{1}{2}\int_{k_1}\!\int_{k_2}\!\int d^dx \,\theta(x^0)e^{i(k_1+k_2)\cdot x}\!\Big\{A_\mu(k_1)A^\mu(k_2)\alpha'(k_2)^{\,2}+\alpha'B(k_1)B(k_2)\notag\\
    & \qquad\qquad\qquad\qquad\qquad\qquad\quad
    +i\alpha'k_2^\mu\left(A_\mu(k_1)B(k_2)-A_\mu(k_2)B(k_1)\,\right)\Big\}\,\notag\\
    &+\,\!\frac{1}{2}\int_{k_1}\!\int_{k_2}\int d^dx \,\delta(x^0)e^{i(k_1+k_2)\cdot x}\nonumber\\
    &\qquad\times\Bigg[i\alpha'(k_1^0+k_2^0)\Big(\alpha'(k_2)^{\,2} A^0(k_1)A^0(k_2)+i\alpha'k_2^0 A^0(k_1)B(k_2)\Big)\nonumber\\ 
&\qquad\qquad+\frac{\theta(-k_1^0)+\theta(k_2^0)}{2}\Big\{i\alpha'(k_1^0-k_2^0)A_\mu(k_1)A^\mu(k_2)\nonumber\\
&\qquad\qquad\qquad\qquad\qquad\qquad-i\alpha'^2(k_1^0-k_2^0)(k_1^0+k_2^0)^2A^0(k_1)A^0(k_2)\notag\\
    &\qquad\qquad\qquad\qquad\qquad\qquad-\alpha'A^0(k_2)B(k_1)\big(1+\alpha'k_1^0(k_1^0+k_2^0)\big)\notag\\
    &\qquad\qquad\qquad\qquad\qquad\qquad+\alpha'A^0(k_1)B(k_2)\big(1+\alpha'k_2^0(k_1^0+k_2^0)\big)\Big\}\Bigg]\,.\label{open massless action}
\end{align}
The terms containing $\delta(x^0)$ contribute to the boundary action, while the terms with $\theta(x^0)$ in \eqref{open massless action} correspond to the bulk action. 
An important point to note is that the contribution to the boundary action in \eqref{open massless action}  is not limited to what we have referred to as the ``boundary'' term in \eqref{new action}, but also includes contributions from the kinetic term.
Such terms arise from
contractions between $\Theta_0[X^0]$ and $\alpha^\mu_n$ in the first term of the action \eqref{new action}. These terms do not contain $\theta(k^0)$, and they vanish as a consequence of energy conservation, $k_1^0+k_2^0=0$.

The bulk part can be written as follows
\footnote{The integration-by-parts term
\begin{align}
 \frac{\alpha'}{2}\int d^dx\,\theta(x^0)\,
 \partial_\rho\left(
 -A_\mu F^{\rho\mu}+A^\rho\mathcal K
 \right)
\end{align}
is included in the boundary action.}
:
\begin{align}
    S_\text{bulk}=&\,\frac{{\alpha'}}{2}\!\int \!d^dx\,\theta(x^0)\Big[
    \frac{1}{2}F_{\mu\nu}F^{\mu\nu}+\mathcal{K}^2
    \Big]\,,\qquad \text{with}\quad\mathcal{K}\equiv B(x)-\partial_\mu A^\mu(x)\,.
\end{align}
This agrees with the result for the bulk part in \cite{Maccaferri:2025onc}. 
The boundary contribution differs from the boundary term in \cite{Maccaferri:2025onc}, since the ``boundary'' condition is not simply of the Dirichlet type. The same situation arises in the closed-string case, and we will discuss it in Section~\ref{sec:boundary condition}. 

Let us check whether the two-point amplitude of the massless string is correctly obtained. The calculation can be carried out in the same way as in the tachyon case, yielding the following result 
\begin{align}
    \mathcal{A}_2=&\,C\times\frac{1}{2}\omega\Big(\Psi_\mu(k_1),\frac{1}{2}\left\{P_+,\left[Q,\Theta_0[X^0]\right]\right\}\Psi_\nu(k_2)\Big)\,+\,(k_1\leftrightarrow k_2)\notag\\
    =&\,C\times\frac{\alpha'}{2i}\eta_{\mu\nu}\,2|k_2^0|\,(2\pi)^{d-1}\delta^{(d-1)}(\mathbf{k_1}+\mathbf{k_2})\,.
\end{align}
In \eqref{open massless action}, imposing the on-shell condition $k^{\,2}=0$ for the fields, and further assuming that we impose the bulk equations of motion on the auxiliary fields—equivalently, that we integrate them out in advance—the remaining terms are all proportional to $k_1^0-k_2^0$. Consequently, as in the tachyon case, after setting $C=\frac{2i}{\alpha'}$ as determined from the tachyon amplitude\eqref{factor C in open}, the desired non-vanishing two-amplitude is obtained if and only if the energy conservation condition $k_1^0+k_2^0=0$ is satisfied.

\subsection{Closed string tachyon}
Next, we extend the construction to the closed string case.
The issue of ghost number deficiency does not arise here, since the symplectic form has been defined with the insertion of $\#$. As a result, the closed string case can be obtained by a straightforward extension of the open string construction. We define the step function $\Theta_0[X^0]$ as follows:
\begin{align}
    \Theta_0[X^0]=\frac{1}{2}\left(\oint_{\mathcal{C}}\!\frac{dz}{2\pi i}\frac{1}{z}-\oint_{\mathcal{C}}\!\frac{d\bar{z}}{2\pi i}\frac{1}{\bar{z}}\right)\left[
    \,\frac{1}{2}\,+\,\int_q \frac{i}{q}e^{i\hat{q}\cdot X}(z,\bar{z})
    \right]\,.
\end{align}
Q is the BRST operator introduced previously in  \eqref{BRST closed}, given by
\begin{align}
    Q=\sum^\infty_{n=-\infty}  
    \left(c_{-n}L^{(\text{m})}_n+\tilde{c}_{-n}\tilde{L}_n^{(\text{m})}\right)
    +
    \!\!\sum^\infty_{\substack{m,n=-\infty\\m\neq n}}
    \!\!\!\frac{(m-n)}{2}
    :c_m c_nb_{-m-n}\,+\,\tilde{c}_m \tilde{c}_n\tilde{b}_{-m-n}\!:
    \,-
    \,(c_0+\tilde{c}_0)\,.
\end{align}
We substitute the tachyon field
\begin{align}
    \Phi=\int_kT(k)c_1\tilde{c}_1\ket{0,k}
\end{align}
into the action and perform the actual calculation. Specifically, the following calculation results are useful:
\begin{align}
    \Theta_0[X^0]\ket{k}=\frac{1}{2}\ket{0,k}\,+\,\int_q \frac{i}{q}\ket{0,k+\hat{q}}\,,
\end{align}
\vspace{-5mm}
\begin{align}
    Qc_1\tilde{c}_1\ket{0,k}=\left(\frac{\alpha'k^2}{4}-1\right)2c_0^+c_1\tilde{c}_1\ket{0,k}\,,
\end{align}
\vspace{-5mm}
\begin{align}
    \left[Q,\Theta_0[X^0]\right]c_1\tilde{c}_1\ket{0,k}=\int_q \frac{i\alpha'}{4}(-2k^0-q)2c_0^+c_1\tilde{c}_1\ket{0,k+\hat{q}}\,.
\end{align}
Here we define $c_0^+ \equiv (c_0 + \tilde{c}_0)/2$. The action then takes the following form:
\begin{align}
     S=& \,\frac{1}{2}\int_{k_1}\!\int_{k_2}T(k_1)T(k_2)\bra{0,k_1}\tilde{c}_{-1}c_{-1}\,(c_0^-)\,\Theta_0[X^0]Qc_1\tilde{c}_1\ket{0,k_2}\notag\\[6pt]
    &\hspace{1cm}+\frac{1}{2}\int_{k_1}\!\int_{k_2}T(k_1)T(k_2)\bra{0,k_1}\tilde{c}_{-1}c_{-1}(c_0^-)\,\frac{1}{2}
      \left\{P_+,\left[Q,\Theta_0[X^0]\right]
\right\}\, c_1\tilde{c}_1\ket{0,k_2}\notag\\[6pt]
      =&\,\frac{1}{2}\int_{k_1}\!\int_{k_2}T(k_1)T(k_2)\Big[
      \!\int d^dx\,\theta(x^0)\Big(\frac{\alpha'k_2^2}{4}-1\Big)e^{i(k_1+k_2)\cdot x}\notag\\[6pt]
      &\hspace{6cm}+i\int d^dx \,\delta(x^0)\frac{\theta(k_2^0)+\theta(-k_1^0)}{2}\frac{\alpha'}{4}(k_1^0-k_2^0)e^{i(k_1+k_2)\cdot x}
      \Big]\,.
\end{align}

As in the open-string tachyon case, the term originating from 
$\{P_+,\big[Q,\Theta_0[X^0]\big]\}$ contains a non-local operator $P_+$ and is therefore not a local boundary contribution in the usual sense, even though it is accompanied by $\delta(x^0)$.  
We postpone this point to Section~\ref{sec:boundary condition} and first verify that the term reproduces the correct two-point amplitude.
We proceed to compute the two-point amplitude of the tachyon, following the same method as in the open string case,
\begin{align}
    \mathcal{A}_2=&
    \,C\times\frac{1}{2}\omega\Big(\Psi(k_1),\frac{1}{2}\left\{P_+,\left[Q,\Theta_0[X^0]\right]\right\}\Psi(k_2)\Big)\,+\,(k_1\leftrightarrow k_2)\notag\\
    =&\,C\times\frac{i\alpha'}{8}\Bigg((k_1^0-k_2^0)\left(\frac{\theta(k_2^0)+\theta(-k_1^0)}{2}-\frac{\theta(k_1^0)+\theta(-k_2^0)}{2}\right)
    \Bigg)(2\pi)^{d-1}\delta^{(d-1)}(\bm{k_1}+\bm{k_2})\notag\\
    =&\,C\times\frac{\alpha'}{8i}\times2|k_2^0|\,(2\pi)^{d-1}\delta^{(d-1)}(\bm{k_1}+\bm{k_2})\,.\label{closed tachyon amp.}
\end{align}
It follows that, in order to reproduce the correct two-point amplitude, the normalization constant $C$ should be chosen as
\begin{align}
    C=\frac{8i}{\alpha'}\,,\label{factor C in closed}
\end{align}
for a closed string.

\subsection{Massless closed string}
In this section, we will describe the construction of the action and two-point amplitude for the massless closed-string sector. The massless closed-string field, including the auxiliary fields, is
\begin{align}
    \Phi = & \int_k -\frac{1}{2}e_{\mu\nu}(k) c_1 \tilde{c}_1 \alpha^\mu_{-1} \tilde{\alpha}^\nu_{-1} |0,k\rangle
    \,+\, \bigl( e(k) c_1 c_{-1} + \bar{e}(k) \tilde{c}_1 \tilde{c}_{-1} \bigr) |0,k\rangle \notag\\[6pt]
    &\hspace{4cm} +\, i\sqrt{\frac{\alpha'}{2}} c_0^+ \left(
    e_\mu(k) c_1 \alpha^\mu_{-1} + \bar{e}_\mu(k) \tilde{c}_1 \tilde{\alpha}^\mu_{-1} \right) |0,k\rangle\,.
\end{align} 
The explicit calculations of the bulk and “boundary” parts obtained by substituting this field into the action \eqref{action section3} are somewhat lengthy. We therefore do not present them here and instead refer the reader to the appendix.
For the moment, we focus on the kinetic term. It can be seen that this corresponds to the classical gravitational action obtained by expanding around the flat metric. 

The $\theta(x^0)$ part of
$\frac12\omega(\Phi,\Theta_0[X^0]Q\Phi)$
can be rewritten as follows:
\begin{align}
    &\frac{1}{2
    }\,\omega(\Phi,\Theta_0[X^0]Q \,\Phi)\notag\\
    &\supset\frac{\alpha'}{8}\!\int_{k_1}\!\int_{k_2}\!\int\!d^dx\,\theta(x^0)e^{i(k_1+k_2)\cdot x}\,\Bigg[
    \frac{k_2^{\,2}}{4}e_{\mu\nu}(k_1)e^{\mu\nu}(k_2)\,-\frac{i}{2}\,e_{\mu\nu}(k_1)e^\mu(k_2)k_2^\nu\,+\frac{i}{2}\,e_{\mu\nu}(k_2)e^\mu(k_1)k_2^\nu\notag\\[6pt]
    &\hspace{3.5cm}+\frac{i}{2}e_{\mu\nu}(k_1)\bar{e}^\nu(k_2)k_2^\mu\,-\frac{i}{2}\,e_{\mu\nu}(k_2)\bar{e}^\nu(k_1)k_2^\mu\,+\,k_2^{\,2}\big(\,e(k_1)\bar{e}(k_2)+e(k_2)\bar{e}(k_1)\,\big)\notag\\[6pt]
    &\hspace{3.5cm}+ie(k_1)\bar{e}_\mu(k_2)k_2^\mu\,-\,ie(k_2)\bar{e}_\mu(k_1)k_2^\mu\,+\,i\bar{e}(k_1)e_\mu(k_2)k_2^\mu\,-\,i\bar{e}(k_2)e_\mu(k_1)k_2^\mu\notag\\[6pt]
    &\hspace{10.5cm}+e_\mu(k_1)e^\mu(k_2)\,+\,\bar{e}^\mu(k_1)\bar{e}_\mu(k_2)\Bigg]\notag\\
    &=\frac{-\alpha'}{8}\int \! d^dx \,\theta(x^0)\times\Bigg[
    \frac{-1}{4}\partial_\rho e_{\mu\nu}\partial^\rho e^{\mu\nu}\,+\,\frac{1}{4}\partial_\rho e_{\mu\nu}\partial^\nu e^{\mu\rho}\,+\,\frac{1}{4}\partial_\rho e_{\nu\mu}\partial^\nu e^{\rho\mu}\notag\\
    &\hspace{5cm}+\Big(\partial_\mu(e-\bar{e})\Big)^2-\frac{1}{2}\partial^\nu\big(e_{\nu\mu}+e_{\mu\nu}\big)\partial^\mu(\bar{e}-e)\,-\,\mathcal{K}_\mu\mathcal{K}^\mu-\bar{\mathcal{K}}_\mu\bar{\mathcal{K}}^\mu
    \Bigg]\,\notag\\
    &\hspace{0.5cm}+\frac{\alpha'}{8}\int d^dx\,\theta(x^0)\,\partial_\rho\Bigg[-\frac{1}{4}\left(
    e^{\mu\nu}\partial^\rho e_{\mu\nu}
    +e^{\mu\rho}\partial^\nu e_{\mu\nu}
    +e^{\rho\nu}\partial^\mu e_{\mu\nu}
    -e_{\mu\nu}\partial^\nu e^{\mu\rho}
    -e_{\mu\nu}\partial^\mu e^{\rho\nu}
    \right)\notag\\
    &\hspace{4.3cm}
    -\frac{1}{2}e_\mu e^{\mu\rho}+\frac{1}{2}\bar{e}_\nu e^{\rho\nu}+e\,\bar{e}^\rho+\bar{e}\,e^\rho+\frac{1}{2}(\bar{e}+e)\partial_\nu (e^{\rho\nu}-e^{\nu\rho})-\partial^\rho(e\bar{e})\,
    \Bigg]\label{Closed action 1st term}
\end{align}
with 
\begin{align}
    \mathcal{K}_\mu\equiv e_\mu\,+\,\frac{1}{2}\partial^\nu e_{\mu\nu}\,-\,\partial_\mu \bar{e}\,,\qquad 
    \bar{\mathcal{K}}_\mu\equiv \bar{e}_\mu\,-\,\frac{1}{2}\partial^\nu e_{\nu\mu}\,-\,\partial_\mu {e}\,.
\end{align}
In the second equality, we perform integration by parts, which generates a surface term. This contribution can be written as a total derivative, and we cannot neglect it. 

Furthermore, by integrating the auxiliary fields $e_\mu$ and $\bar{e}_\mu$, setting $\mathcal{K}_\mu$ and $\bar{\mathcal{K}}_\mu$ to zero, decomposing $e_{\mu\nu}$ into its symmetric and antisymmetric components, and introducing a field redefinition for the dilaton $\phi$, 
\begin{align}
    e_{\mu\nu}&=b_{\mu\nu}+h_{\mu\nu}\,,\qquad b_{\mu\nu}=-b_{\nu\mu}\,,\qquad h_{\mu\nu}=h_{\nu\mu},\\[6pt]
    &\hspace{2cm}
    \phi=\frac{1}{4}(h^\mu_{\,\,\mu}-2\bar{e}+2e),\\[6pt]
    &\hspace{1.5cm}H_{\mu\nu\rho}=\partial_\mu b_{\nu\rho}+\partial_\nu b_{\rho\mu}+\partial_\rho b_{\mu\nu}\,.
\end{align}
The bulk action can be rewritten in the following form
\begin{align}
    S_\text{bulk}&=\frac{\alpha'}{8}\int d^dx\,\theta(x^0)\left(\frac{1}{12}H_{\mu\nu\rho}H^{\mu\nu\rho}\right)\notag\\
    &\hspace{0.5cm}+\frac{\alpha'}{8}\int d^dx\,\theta(x^0)\Bigg[
    \frac{1}{4}\partial_\rho h_{\mu\nu}\partial^\rho h^{\mu\nu}-\frac{1}{2}\partial_\rho h_{\mu\nu}\partial^\nu h^{\mu\rho}+\frac{1}{4}\partial^\mu(2h_{\mu\nu}-\eta_{\mu\nu}h^\rho_{\,\,\rho})\partial^\nu h^\lambda_{\,\,\lambda}\notag\\
    &\hspace{8cm}-2\partial^\mu(h_{\mu\nu}-\eta_{\mu\nu}h^\rho_{\,\,\rho})\partial^\nu \phi-4\partial_\mu \phi \partial^\mu \phi
    \Bigg]\,.
\end{align}
This corresponds, up to an overall normalization, to the quadratic
expansion of the following low-energy effective action:
\begin{align}
    S_\text{bulk}=\int_Md^dx \,\theta(x^0) \sqrt{-g}e^{-2\phi}\left(R\,-\,\tfrac{1}{12}H^2\,+\,4(\nabla \phi)^2\right)\,,
\end{align}
where $R$ is the Ricci scalar. This agrees with the result for the bulk part in \cite{Maccaferri:2025onc}. 

It receives contributions from integration by parts,
from the $\delta(x^0)$ part of
$\frac12\omega(\Phi,\Theta_0[X^0]Q\Phi)$,
and from $\frac12\omega(\Phi,\mathcal B\Phi)$.
Combining these contributions, we obtain
\begin{align}
     S_\text{bdy}&=\frac{\alpha'}{8}\int d^dx\,\theta(x^0)\,\partial_\rho\Big[-\frac{1}{4}\left(
    e^{\mu\nu}\partial^\rho e_{\mu\nu}
    +e^{\mu\rho}\partial^\nu e_{\mu\nu}
    +e^{\rho\nu}\partial^\mu e_{\mu\nu}
    -e_{\mu\nu}\partial^\nu e^{\mu\rho}
    -e_{\mu\nu}\partial^\mu e^{\rho\nu}
    \right)\notag\\
    &\hspace{3.8cm}
    -\frac{1}{2}e_\mu e^{\mu\rho}+\frac{1}{2}\bar{e}_\nu e^{\rho\nu}+e\,\bar{e}^\rho+\bar{e}\,e^\rho+\frac{1}{2}(\bar{e}+e)\partial_\nu (e^{\rho\nu}-e^{\nu\rho})-\partial^\rho(e\bar{e})\,
    \Big]\notag
\end{align}
\vspace{-0.8cm}
\begin{align}
&\quad+\frac{\alpha'}{8}
\int_{k_1}\!\int_{k_2}\!\int d^dx\,
\delta(x^0)e^{i(k_1+k_2)\cdot x}
\frac{\theta(k_2^0)+\theta(-k_1^0)}{2}
\notag\\
&\qquad\times\Bigg[
\frac{i}{4}(k_1^0-k_2^0)
\Bigg(
\eta^{\mu\rho}\eta^{\nu\lambda}
-\frac{\alpha'(k_1^0+k_2^0)^2}{2}
 \eta^{\mu\nu\rho\lambda}
+\frac{\alpha'^2(k_1^0+k_2^0)^4}{4}
 \eta^{\mu0}\eta^{\nu0}\eta^{\rho0}\eta^{\lambda0}
\Bigg)
e_{\rho\lambda}(k_1)e_{\mu\nu}(k_2)
\notag\\[6pt]
&\hspace{3.3cm}
-\frac12
\left(1+\frac{\alpha'k_2^0(k_1^0+k_2^0)}{2}\right)
\left(
\eta^{\mu\rho}\eta^{\lambda0}
-\frac{\alpha'(k_1^0+k_2^0)^2}{2}
 \eta^{\mu0}\eta^{\rho0}\eta^{\lambda0}
\right)
e_{\rho\lambda}(k_1)e_\mu(k_2)
\notag\\[6pt]
&\hspace{3.3cm}
+\frac12
\left(1+\frac{\alpha'k_2^0(k_1^0+k_2^0)}{2}\right)
\left(
\eta^{\mu\lambda}\eta^{\rho0}
-\frac{\alpha'(k_1^0+k_2^0)^2}{2}
 \eta^{\mu0}\eta^{\lambda0}\eta^{\rho0}
\right)
e_{\rho\lambda}(k_1)\bar e_\mu(k_2)
\notag\\[6pt]
&\hspace{3.3cm}
+\frac12
\left(1+\frac{\alpha'k_1^0(k_1^0+k_2^0)}{2}\right)
\left(
\eta^{\mu\rho}\eta^{\nu0}
-\frac{\alpha'(k_1^0+k_2^0)^2}{2}
 \eta^{\mu0}\eta^{\rho0}\eta^{\nu0}
\right)
e_{\mu\nu}(k_2)e_\rho(k_1)
\notag\\[6pt]
&\hspace{3.3cm}
-\frac12
\left(1+\frac{\alpha'k_1^0(k_1^0+k_2^0)}{2}\right)
\left(
\eta^{\nu\rho}\eta^{\mu0}
-\frac{\alpha'(k_1^0+k_2^0)^2}{2}
 \eta^{\nu0}\eta^{\rho0}\eta^{\mu0}
\right)
e_{\mu\nu}(k_2)\bar e_\rho(k_1)
\notag\\[6pt]
&\hspace{10cm}
+i(k_1^0-k_2^0)
\bigl(
e(k_1)\bar e(k_2)+e(k_2)\bar e(k_1)
\bigr)
\notag\\[6pt]
&\hspace{6.5cm}
+\left(1+\frac{\alpha'k_1^0(k_1^0+k_2^0)}{2}\right)
\eta^{\mu0}
\bigl(
e(k_1)\bar e_\mu(k_2)+\bar e(k_1)e_\mu(k_2)
\bigr)
\notag\\[6pt]
&\hspace{6.5cm}
-\left(1+\frac{\alpha'k_2^0(k_1^0+k_2^0)}{2}\right)
\eta^{\nu0}
\bigl(
e(k_2)\bar e_\nu(k_1)+\bar e(k_2)e_\nu(k_1)
\bigr)
\Bigg]\notag
\end{align}
\vspace{-0.8cm}
%%%%%%%%%%%%%%%  O(α'^2)
\begin{align}
&\hspace{0.3cm}+\frac{i\alpha'^2}{16}
\int_{k_1}\!\int_{k_2}\!\int d^dx\,
\delta(x^0)e^{i(k_1+k_2)\cdot x}
(k_1^0+k_2^0)\times\Bigg[\,\,
\frac{k_2^2}{4}\eta^{\mu\nu\rho\lambda}
e_{\rho\lambda}(k_1)e_{\mu\nu}(k_2)
\notag\\
&\hspace{1.4cm}-\frac{i}{2}\eta^{\mu\nu\rho\lambda}
e_{\rho\lambda}(k_1)e_\mu(k_2)(k_2)_\nu
+\frac{i}{2}e^{0\nu}(k_2)e^0(k_1)(k_2)_\nu
+\frac{i}{2}\eta^{\mu\nu\rho\lambda}
e_{\rho\lambda}(k_1)\bar e_\nu(k_2)(k_2)_\mu
\notag\\
&\hspace{1.4cm}
-\frac{i}{2}(k_2)_\mu e^{\mu0}(k_2)\bar e^0(k_1)
-ik_2^0\bigl(
\bar e^0(k_1)e(k_2)+e^0(k_1)\bar e(k_2)
\bigr)
+e^0(k_1)e^0(k_2)
+\bar e^0(k_1)\bar e^0(k_2)
\Bigg]
\notag\\[6pt]
&\quad-\frac{i\alpha'^3}{32}
\int_{k_1}\!\int_{k_2}\!\int d^dx\,
\delta(x^0)e^{i(k_1+k_2)\cdot x}
(k_1^0+k_2^0)^3 e^{00}(k_1)
\times\left[
\frac{k_2^2}{4}e^{00}(k_2)
-\frac{i}{2}k_2^0e^0(k_2)
+\frac{i}{2}k_2^0\bar e^0(k_2)
\right]\,.
\label{massless closed boundary}
\end{align}
Although the expression is somewhat lengthy, its structure is simple. 
There are two qualitatively different types of contributions. 
The terms produced by integrating by parts are genuinely localized at the boundary $x^0=0$. 
By contrast, the contributions arising from \eqref{Closed action 2nd term} contain the step functions of energy, such as $\theta(k_2^0)$ and $\theta(-k_1^0)$, and are therefore non-local in the time direction. 
Thus, despite the presence of $\delta(x^0)$, these terms should not be regarded as local boundary terms in the usual sense. Such terms also arise in the derivation of the equations of motion. Owing to this non-locality, the boundary conditions are not of the standard type. We discuss this point in the next section.

Imposing the on-shell condition on the fields, we again find that the amplitude is non-vanishing only when $k_1^0+k_2^0=0$, and it takes the following form:
\begin{align} 
    \mathcal{A}_2=C\times\frac{\alpha'}{8i} \times \eta_{\mu\rho}\eta_{\nu\lambda}\,2|k_2^{\,0}|\,(2\pi)^{d-1}\delta^{(d-1)}(\bm{k_1}+\bm{k_2})\,.\label{closed massless amp.}
\end{align}
Upon using the factor $C=\frac{8i}{\alpha'}$, determined from the closed-string tachyon amplitude\eqref{factor C in closed}, the desired two-point amplitude is obtained.

\subsection{Comment on the ``boundary'' terms and the ``boundary'' conditions}\label{sec:boundary condition}

Finally, let us comment on the “boundary” terms appearing in the above examples and on the “boundary” conditions consistent with them.
As can be seen from the above examples, there are two types of terms other than the standard bulk terms appearing in the action. 
The first type consists of terms proportional to the factor
\begin{align}
    \frac{\theta(k_2^0)+\theta(-k_1^0)}{2}\,.\label{the factor}
\end{align}
Although this term also contains a factor of $\delta(x^0)$, it is not localized on the boundary $x^0=0$. Let us see this explicitly in the case of the open-string tachyon:
\begin{align}
    \frac{i\alpha'}{2}\int_{k_1}\int_{k_2}T(k_1)T(k_2)\int \dd^dx\,\delta(x_0) \frac{\theta(k_2^0)+\theta(-k_1^0)}{2}(k_1^0-k_2^0) e^{i(k_1+k_2)\cdot x}\,.\label{tachyon ``boundary'' momemtum}
\end{align}
Using the integral representation of the step function
\begin{align}
    \theta(k^0)=\int_{-\infty}^\infty \frac{\dd y^0}{2\pi}\frac{i}{y^0} e^{-ik^0y^0} +\frac{1}{2}\,,
\end{align}
and then performing the integrations
over $k_1$ and $k_2$, this term can be written as follows
\begin{align}
    \frac{-i\alpha'}{2}\int \dd^{d-1} \vb{x}\,\int_{-\infty}^\infty \frac{\dd y^0}{2\pi}\frac{1}{y^0}\Big(\partial_0T(0;\vb{x})T(y^0;\vb{x})-T(0;\vb{x})\partial_0T(y^0;\vb{x})\Big)\,.\label{tachyon ``boundary''}
\end{align}
Here, $T(x^0;\vb{x})$ denotes the 
coordinate-space field obtained by Fourier transforming $T(k)$, where $x^0$ is the time component and $\vb{x}$ denotes the spatial components.
To avoid possible confusion, we note that $\partial_0 T(0;\vb{x})$ should be understood as $\partial_t T(t;\vb{x})|_{t=0}$.
This term is integrated over $y^0$ from $-\infty$ to $+\infty$, and hence is not localized.
Terms of this kind arise from the second term in the action \eqref{new action}, which is referred to in this paper as the “boundary” term.
Since, as seen above, such terms are not actually localized on the boundary, we use quotation marks and call them “boundary” terms.

The other type consists of terms that are not proportional to the factor \eqref{the factor}, and are genuinely localized on the boundary.
Such terms arise from contractions between $\Theta_0[X^0]$ and $\alpha_n^\nu$ or $\tilde{\alpha}_n^\mu$ in the first term of the action \eqref{new action}.
As is clear from their origin, such terms are proportional to the equations of motion and therefore do not contribute on shell.
Moreover, the contributions arising from this contraction are subleading in $\alpha'$, and are regarded as genuinely stringy contributions.
Furthermore, these terms are proportional to $k_1^0+k_2^0$, or equivalently, are total derivative terms with respect to $x^0$.
Thus, they do not contribute when energy is conserved.
In the tachyon case, such terms do not arise, since there is no oscillator $\alpha_n^\mu$ or $\tilde{\alpha}_n^\mu$ that can contract with $\Theta_0[X^0]$.

Next, let us consider the “boundary” conditions compatible with these terms.
That is, we seek conditions under which the variation of the action 
\begin{align}
    \delta S=
\omega \!\left(
\,\delta\Phi,\,\Theta_0[X^0] Q\,\Phi
\right)
+
\frac{1}{2}\omega\big(\
\delta\Phi,
\left\{
P_+,\left[Q,\Theta_0[X^0]\right]
\right\}\,\Phi
\big)\label{variation}
\end{align}
vanishes.
The first term yields the equations of motion, but there are also contributions arising from the boundary terms of the latter type classified above.
However, these terms vanish once the equations of motion are imposed, and therefore do not lead to any non-trivial conditions.
Therefore, it suffices to consider appropriate conditions under which the second term in \eqref{variation} vanishes.
Note that this term is of the former type discussed above, i.e., a non-local term.
Thus, rather than boundary conditions in the usual sense, what should be specified is an appropriate class of functions for which the corresponding variations vanish.
Since $\delta\Phi$ appears in the first slot of the symplectic form $\omega$, this term may at first sight seem to vanish under the Dirichlet boundary condition. However, this is not the case.
This is because, at the level of zero modes, $\big[Q,\Theta_0[X^0]\big]$ is an operator containing both $\delta(x^0)\partial_0$ and $\partial_0\delta(x^0)$.
The former contribution is proportional to the component field $\delta\phi_i$ of $\delta\Phi$, and hence vanishes under the Dirichlet boundary condition.
In contrast, the latter contains terms proportional to $\partial_0\delta\phi_i$, which do not vanish under the Dirichlet condition; such terms are instead typically eliminated by the Neumann boundary condition.
For the choice \eqref{P_pm}, the projected fields at $x^0=0$ depend on the full time dependence of the fields off shell. The mode prescription considered below therefore cannot, by itself, be specified using initial-surface data alone.

Intuitively, the presence of $P_+$ suggests that the appropriate conditions should be imposed on the energy modes.
Let us examine this point more explicitly for the open-string tachyon. We take both $T$ and its variation $\delta T$ to be real:
\begin{align}
          i\alpha'\int_{k_1}\int_{k_2}\delta T(k_1)T(k_2)\int \dd^dx\,\delta(x_0) \frac{\theta(k_2^0)+\theta(-k_1^0)}{2}(k_1^0-k_2^0) e^{i(k_1+k_2)\cdot x}\,.\label{tachyon boundary deviation}
\end{align}
Since we have scattering processes in mind, we now impose the on-shell conditions on both $\delta T(k_1)$ and $T(k_2)$.
As discussed in Section~\ref{sec:``Boundary term'' produces two point amplitude}, energy is then conserved, so that $k_1^0+k_2^0=0$.
Under this condition, we can rewrite
\begin{align}
\frac{\theta(k_2^0)+\theta(-k_1^0)}{2}(k_1^0-k_2^0)
&=
-2\theta(-k_1^0)\theta(k_2^0)\,k_2^0 \,.
\end{align}
Here, $\theta(k_2^0)$ and $\theta(-k_1^0)\theta(k_2^0)$ differ at $k_1^0=k_2^0=0$, but this difference does not contribute because of the factor $k_2^0$.
Substituting this expression into \eqref{tachyon boundary deviation} and transforming to coordinate space, we obtain
\begin{align}
    2\alpha'\int\dd^{d-1} \vb{x}\,\delta T_-(0;\vb{x})\partial_0T_+(0;\vb{x})\,,
\end{align}
where $T_\pm(x^0;\vb{x})$ are defined by
\begin{align}
    T_\pm(x^0;\vb{x}):=\int_k \theta(\pm k^0) T(k) e^{ik\cdot x}\,.
\end{align}
Since the field and its variation are real, $T_-=T_+^*$ and $\delta T_-=(\delta T_+)^*$.
We impose $\delta T_-(0;\vb{x})=0$, which fixes the incoming mode data and their reality-related conjugates without requiring the field itself to vanish.\footnote{For the nonzero-energy on-shell modes considered here, imposing $T_+=0$ or $\partial_0T_+=0$ at the initial surface would force $T=\partial_0T=0$ there and exclude nontrivial incoming data. We therefore constrain the variation.}
Using the energy-conservation condition $k_1^0+k_2^0=0$, the same term can also be written as
\begin{align}
-2\alpha'\int\dd^{d-1} \vb{x}\,\partial_0\delta T_-(0;\vb{x})T_+(0;\vb{x})\,.
\end{align}
The condition $\partial_0\delta T_-(0;\vb{x})=0$ also eliminates this expression.
For nonzero-energy on-shell modes it is equivalent to $\delta T_-(0;\vb{x})=0$, on which we focus below.

At first sight, this prescription appears to admit an interpretation as an initial condition for a scattering problem.
However, the above argument relies on the on-shell conditions for both $T$ and $\delta T$, together with the energy conservation derived from these conditions.
These on-shell conditions require $T$ and $\delta T$ to satisfy the equations of motion for all times.
Requiring the equations of motion to hold only at $x^0=0$ is insufficient.
Indeed, the condition
\begin{align*}
0
&=\left.
\left(\partial_0^2-\nabla^2+m^2\right)T(x)
\right|_{x^0=0}
\\
&=\int_k
\left[-(k^0)^2+\bm{k}^2+m^2\right]
T(k)e^{i\bm{k}\cdot\vb{x}}
\end{align*}
does not imply that $T(k)$ has support only on
$k^0=\pm\sqrt{\bm{k}^2+m^2}$, since contributions from different temporal frequencies can cancel in the integral.
The same observation applies to $\delta T$.
Thus, imposing the equations of motion only at $x^0=0$ does not imply the energy conservation used above, and the above argument does not establish a variational prescription based solely on initial data.

To address this issue, in the next section we consider an alternative off-shell choice of $P_\pm$ that is local in time.
With this choice, the ``boundary'' term becomes a genuine boundary term supported at $x^0=0$, and the boundary conditions that eliminate the boundary contribution to the variation can be interpreted as initial conditions.

\section{From a ``boundary'' term to a boundary term}
\label{sec:time-local}\color{black}

The construction in Section~\ref{sec:Restoring cyclicity}
leaves freedom in the off-shell choice of the energy operators.
We introduce new operators at each mass level with mass squared $m^2$\footnote{
The square roots and their inverses, including those in $\mathcal{K}$,
are defined using the branch $\sqrt{\bm{k}^2+m^2+i0}$, which also
covers $\bm{k}^2+m^2<0$.
The zero set $\bm{k}^2+m^2=0$ has measure zero, and the inverse
square-root singularities are locally integrable in $d=26$.
We use distributional Fourier kernels for regular momentum-space
wave packets, leaving modes supported entirely on this zero set aside.
} by
\begin{align}
    P_\pm^{(\mathrm{new})}\Phi(k)
    =\frac12\left(
        1\pm\frac{k^0}{\sqrt{\bm{k}^2+m^2}}
    \right)\Phi(k).
    \label{sec5:projectors}
\end{align}
We rename the original operators in \eqref{P_pm} as $P_\pm^{(\mathrm{old})}:=\theta(\pm p^0)$.
Like $P_\pm^{(\mathrm{old})}$, the new operators satisfy
\begin{align}
    P_+^{(\mathrm{new})}+P_-^{(\mathrm{new})}&=\mathbb{I},
    \label{sec5:sum}\\
    \omega(P_+^{(\mathrm{new})}\Phi_1,\Phi_2)
    &=\omega(\Phi_1,P_-^{(\mathrm{new})}\Phi_2).
    \label{sec5:bpz}
\end{align}
The second identity follows since BPZ-paired states have
opposite momenta and the same level, and $\bm{k}^2$ is unchanged.
With this replacement, $\mathcal{B}$ in \eqref{proposal B} satisfies
\begin{align}
    \mathcal{B}+\mathcal{B}^*
    &=\frac12
      \big\{P_+^{(\mathrm{new})}+P_-^{(\mathrm{new})},\left[Q,\Theta_0[X^0]\right]\big\}
      =\left[Q,\Theta_0[X^0]\right],
    \label{sec5:cyclicity}
\end{align}
and cyclicity is restored without imposing the equations of motion.
The choice \eqref{sec5:projectors} is distinguished by
reproducing the two-point amplitude while being local in time,
as shown below.

On shell, for external states with the equations of motion imposed for all times, $k^0=\pm\sqrt{\bm{k}^2+m^2}$, and hence
\begin{align}
    P_\pm^{(\mathrm{new})}\Phi(k)
    =P_\pm^{(\mathrm{old})}\Phi(k)
    =\theta(\pm k^0)\Phi(k).
    \label{sec5:on-shell-projectors}
\end{align}
In \eqref{two pt amp boundary}, the BPZ relation \eqref{sec5:bpz} moves the leftmost $P_+^{(\mathrm{new})}$ onto the first external field as $P_-^{(\mathrm{new})}$.
Both operators then act on on-shell fields, so all the two-point amplitude calculations remain unchanged.
We check this explicitly for the open-string tachyon below.

With the Fourier convention
$e^{ik\cdot x}=e^{-ik^0x^0+i\bm{k}\cdot\bm{x}}$,
the new operators can be written as
\begin{align}
    P_\pm^{(\mathrm{new})}
    =\frac12\left(
        1\pm\frac{i\partial_0}{\sqrt{-\nabla^2+m^2}}
    \right).
    \label{sec5:local-operator}
\end{align}
The new operators are local in time, depending only on the field and its first time derivative at the same time, but remain non-local in space.
Below, we write $T_\pm:=P_\pm^{(\mathrm{new})}T$.

For the open-string tachyon, $m^2=-1/\alpha'$.
We replace $P_+^{(\mathrm{old})}$ in the additional term of
\eqref{open tachyon action} by \eqref{sec5:projectors}.
The contribution from the constant part of $P_+^{(\mathrm{new})}$
vanishes under interchange of $k_1$ and $k_2$.
Since $\Theta_0[X^0]$ shifts only $k^0$, the two factors
of $P_+^{(\mathrm{new})}$ in the anticommutator have the same square-root
denominator, and the remaining energy factors combine into
$(k_1^0-k_2^0)^2$, giving
\begin{align}
    S_{\mathrm{bdy}}
    =\frac{-i\alpha'}8
    \int_{k_1}\int_{k_2}
    (2\pi)^{d-1}
    \delta^{(d-1)}(\bm{k}_1+\bm{k}_2)
    \frac{(k_1^0-k_2^0)^2}
         {\sqrt{\bm{k}_1^2-1/\alpha'}}
    T(k_1)T(k_2).
    \label{sec5:boundary-momentum}
\end{align}
Using the Fourier convention above, this becomes
\begin{align}
    S_{\mathrm{bdy}}
    =\frac{-i\alpha'}4
    \int d^{d-1}\bm{x}\,
    \left[
        \partial_0T\,
        \frac{1}{\sqrt{-\nabla^2-1/\alpha'}}\,
        \partial_0T
        -
        T\,
        \frac{1}{\sqrt{-\nabla^2-1/\alpha'}}\,
        \partial_0^2T
    \right]_{x^0=0}.
    \label{sec5:boundary-position}
\end{align}
Here the inverse square root acts only on the spatial coordinates,
and $\partial_0^2T$ appears since $\big[Q,\Theta_0[X^0]\big]$ contributes one
time derivative in addition to that in $P_\pm^{(\mathrm{new})}$.

To make the spatial non-locality explicit, introduce the kernel 
\begin{equation}
    \mathcal{K}(\bm{r}):=
    \int
    \frac{d^{d-1}\bm{p}}{(2\pi)^{d-1}}\,
    \frac{e^{i\bm{p}\cdot\bm{r}}}
         {\sqrt{\bm{p}^{\,2}-1/\alpha'}}\,.
    \label{sec5:spatial-kernel}
\end{equation}
The Fourier integral is understood in the distributional sense.
For a test function $f$,
\begin{align}
    \left[
        \frac{1}{\sqrt{-\nabla^2-1/\alpha'}}\,f
    \right](x^0,\bm{x})
    =
    \int d^{d-1}\bm{y}\,
    \mathcal{K}(\bm{x}-\bm{y})\,f(x^0,\bm{y})\,.
    \label{sec5:inverse-square-root-kernel}
\end{align}
The boundary contribution
\eqref{sec5:boundary-position} can therefore be written as 
\begin{align}
    S_{\mathrm{bdy}}
    &=
    \frac{-i\alpha'}4
    \int d^{d-1}\bm{x}\,d^{d-1}\bm{y}\,
    \mathcal{K}(\bm{x}-\bm{y})
    \notag\\
    &\qquad\times
    \Big[
        \partial_0 T(x^0,\bm{x})\,
        \partial_0 T(x^0,\bm{y})
        -
        T(x^0,\bm{x})\,
        \partial_0^2 T(x^0,\bm{y})
    \Big]_{x^0=0}\,.
    \label{sec5:boundary-position-kernel}
\end{align}
This expression couples fields at distinct spatial points on the same time slice $x^0=0$, making the spatial non-locality manifest.

The additional term is therefore a genuine boundary term supported at $x^0=0$ even off shell, although it remains non-local in space.
On shell, using $\partial_0^2T=(\nabla^2+1/\alpha')T$, it becomes
$-i\alpha'\int d^{d-1}\bm{x}\,T_+\sqrt{-\nabla^2-1/\alpha'}\,T_-$,
which is independent of the time slice when the equations of motion hold for all times.

To compute the two-point amplitude, we use
$C=2i/\alpha'$ as in \eqref{factor C in open}
and $\delta T(p)/\delta T(k)=(2\pi)^d\delta^{(d)}(p-k)$.
Before imposing the on-shell conditions, we obtain
\begin{align}
    \frac{2i}{\alpha'}
    \frac{\delta^2 S_{\mathrm{bdy}}}
         {\delta T(k_1)\delta T(k_2)}
    =
    \frac{(k_1^0-k_2^0)^2}
         {2\sqrt{\bm{k}_1^2-1/\alpha'}}
    (2\pi)^{d-1}
    \delta^{(d-1)}(\bm{k}_1+\bm{k}_2).
    \label{sec5:two-point-kernel}
\end{align}
On shell, $(k_i^0)^2=\bm{k}_i^2-1/\alpha'$ together with
spatial momentum conservation gives $k_1^0=\pm k_2^0$.
The case $k_1^0=k_2^0$ vanishes because of
$(k_1^0-k_2^0)^2$, so energy conservation is not imposed
by hand.
For $k_1^0=-k_2^0$, $\sqrt{\bm{k}_1^2-1/\alpha'}=|k_2^0|$,
and hence
\begin{align}
    \mathcal{A}_2
    =2|k_2^0|(2\pi)^{d-1}
      \delta^{(d-1)}(\bm{k}_1+\bm{k}_2),
    \label{sec5:amplitude}
\end{align}
in agreement with \eqref{open tachyon amp}.

The second term of \eqref{variation} becomes
\begin{align}
    &\frac12\omega\big(\delta\Phi,
    \{P_+^{(\mathrm{new})},[Q,\Theta_0[X^0]]\}\Phi\big)
    \notag\\*
    &=-\frac{\alpha'}2\int d^{d-1}\bm{x}\,
    \big[W(\delta T,T_+)+W(\delta T_-,T)\big]_{x^0=0},
    \label{sec5:variation}
\end{align}
where $W(A,B):=\partial_0A\,B-A\,\partial_0B$.
Here we define $\delta T_\pm:=P_\pm^{(\mathrm{new})}\delta T$.

As in Section~\ref{sec:boundary condition}, we take $T$ and $\delta T$ to be real.
For the variational problem, we impose the equations of motion only at the initial surface:
\begin{align}
    \left.(\partial_0^2-\nabla^2-1/\alpha')T\right|_{x^0=0}=0,
    \qquad
    \left.(\partial_0^2-\nabla^2-1/\alpha')\delta T\right|_{x^0=0}=0.
    \label{sec5:boundary-eom}
\end{align}
The second condition is the variation of the first.
These imply $P_\pm^{(\mathrm{new})}T_\mp=0$ and
$P_\pm^{(\mathrm{new})}\delta T_\mp=0$ at $x^0=0$.
Integrating by parts in the spatial coordinates to transfer $\nabla^2$ between the factors, \eqref{sec5:variation} reduces to
\begin{align}
    -2i\alpha'\int d^{d-1}\bm{x}\,
      \delta T_-(0;\bm{x})\sqrt{-\nabla^2-1/\alpha'}\,T_+(0;\bm{x}).
    \label{sec5:reduced-variation}
\end{align}
Thus \eqref{sec5:boundary-eom}, together with
\begin{align}
    \delta T_-(0;\bm{x})=0,
    \label{sec5:initial-alternatives}
\end{align}
eliminates the boundary variation.
This has the same form as the variation condition in Section~\ref{sec:boundary condition}, now using $P_\pm^{(\mathrm{new})}$.

The condition fixes the initial value of $T_-$, which specifies
the incoming data for scattering states.
Here, on-shell boundary data means compatibility with the equations of motion at the initial surface, not mass-shell support of the full temporal Fourier transform.
The fields and variations remain unrestricted by the equations of motion away from this surface, so this is an initial-data prescription.

\section{Discussion}\label{sec:Discussion}

In this paper, we proposed a quadratic action for free string field theory in the presence of a spacelike interface at $x^0=0$,
\begin{align}
    S=\frac{1}{2}\omega\big(\Phi,\Theta_0[X^0]Q\Phi\big)
    +\frac{1}{4}\omega\Big(\Phi,\big\{P_+,\big[Q,\Theta_0[X^0]\big]\big\}\Phi\Big)\,,
\end{align}
and studied its basic properties in both open and closed bosonic string field theory.
The choice $P_\pm=P_\pm^{(\mathrm{old})}$ introduces non-locality in time, whereas $P_\pm=P_\pm^{(\mathrm{new})}$ is local in time and non-local in space. For the open-string tachyon, the latter choice gives a term supported at $x^0=0$.
The second term is required because the insertion of $\Theta_0[X^0]$ in the first term spoils the cyclicity of the symplectic form.
With this additional contribution included, cyclicity is restored for both choices. For the time-local choice, we have also identified an initial-data prescription that eliminates the boundary variation in the open-string tachyon sector.
We also showed that, when evaluated on on-shell states, the second term reproduces the correct tree-level two-point amplitude, in agreement with the operator-formalism analysis based on the mostly BRST-exact operator.
This is reminiscent of the relative-$L_\infty$ perspective, where the generalised minimal model of a relative $L_\infty$ algebra captures the full tree-level $S$-matrix, including the otherwise missing two-point contribution \cite{Alfonsi:2024utl}.

A characteristic feature of the present construction is that the additional contribution naturally takes the form
$\{P_+,\big[Q,\Theta_0[X^0]\big]\}$.
Here $\big[Q,\Theta_0[X^0]\big]$ represents the insertion of the mostly BRST-exact operator, while $\{P_+,\,\cdot \,\}$ removes the extra sign factor that appears in the operator-formalism computation.
As a result, this combination reproduces the correct two-point amplitude in a direct way.
In this sense, $\{P_+,\big[Q,\Theta_0[X^0]\big]\}$ provides a string-field-theoretic analogue of the relative $L_\infty$ mechanism by which an additional boundary contribution completes the tree-level \(S\)-matrix.

The two choices of $P_\pm$ give the same on-shell two-point amplitudes but define different off-shell actions.
For $P_\pm^{(\mathrm{old})}$, the additional term is non-local in time and depends on the full time dependence of a general off-shell field.
We use quotation marks to indicate this feature of the construction in Sections~\ref{sec:preliminaries}--\ref{sec:example}.
For $P_\pm^{(\mathrm{new})}$, the open-string tachyon term is a genuine contribution supported at $x^0=0$, although it remains spatially non-local.
The two-point amplitude therefore does not determine the off-shell variational prescription.

The variational analysis makes the distinction between the two choices explicit.
For $P_\pm^{(\mathrm{old})}$, the argument in Section~\ref{sec:boundary condition} uses globally on-shell $T$ and $\delta T$, together with the resulting energy conservation.
Imposing the equations of motion only at $x^0=0$ does not justify that argument.
For $P_\pm^{(\mathrm{new})}$, the boundary equations \eqref{sec5:boundary-eom} instead reduce the boundary variation to \eqref{sec5:reduced-variation}, which vanishes when $\delta T_-(0;\bm{x})=0$.
The allowed variations preserve these boundary conditions but need not satisfy the equations of motion away from the initial surface.
Fixing $T_-(0;\bm{x})$ specifies a component of the initial data
without requiring the field to vanish.
The on-shell condition on the boundary is a compatibility condition on the initial data, rather than a global on-shell restriction.
This initial-data prescription has been established explicitly for the open-string tachyon; extending it to the other component fields remains to be investigated.

There are several natural directions for future work.
First, the present analysis concerns only the linear part of the homotopy-algebraic structure, namely the differential $Q$.
It is therefore natural to ask for a non-linear extension including the higher products that encode interactions.
In such a framework, however, the presence of a spacelike interface or spacetime boundary is expected to require more intricate non-local ``boundary'' terms in order to restore cyclicity beyond the quadratic level.
Understanding how these terms are organized would be essential for formulating interacting string field theory in the presence of boundaries or interfaces.

Second, a parallel extension to superstring field theory should also be possible.
Since the mostly BRST-exact operator already plays an important role in operator-formalism analyses of two-point amplitudes, one may expect a related structure to arise there as well.
At the same time, the presence of picture changing and related subtleties suggests that the superstring version should exhibit a richer structure.
Understanding this extension may also clarify the relation between non-local ``boundary'' contributions, cyclicity, and physical observables in superstring field theory.

Third, it would also be interesting to clarify the relation to the covariant phase space formalism described in terms of homotopy algebras \cite{Bernardes:2025uzg, Bernardes:2025zzu, Bernardes:2025zkj, Bernardes:2026egp}.
For a general $\mathcal{B}$ satisfying \eqref{def of B}, not necessarily the specific choice proposed in \eqref{proposal B}, let us consider the two-form constructed from the “boundary” term,
\begin{align}
    \omega(\dd\Phi,\,\mathcal{B}\dd\Phi)\,.
\end{align}
Here, $\dd\Phi$ is a one-form, and two factors of $\dd\Phi$ acquire a minus sign when they are interchanged.
This term can be written as
\begin{align}
    \frac{1}{2}\,\omega\big(\dd\Phi,\,(\mathcal{B}+\mathcal{B}^*)\dd\Phi\big)
=
\frac{1}{2}\,\omega\big(\dd\Phi,\,\big[Q,\Theta_0[X^0]\big]\,\dd\Phi\big)\,.
\end{align}
This agrees with the symplectic form proposed in \cite{Bernardes:2025uzg} in the absence of interactions, when the sigmoid $\sigma$ is chosen to be $\Theta_0[X^0]$.
In \cite{Bernardes:2025uzg}, the symplectic form was originally introduced as an object satisfying suitable conditions.
However, in the covariant phase space formalism, the symplectic form is in principle constructed from the boundary term of the action.
Therefore, this agreement is natural.

A derivation of this symplectic form from the boundary contribution
to the variation of an $L_\infty$ action was also given by
Ali and Stettinger \cite{Ali:2026BEF}.
In the covariant phase space formalism, the fields satisfy the
equations of motion, while their tangent variations satisfy the
linearized equations.
This is consistent with our result in Section~\ref{sec:time-local},
where the boundary conditions require the open-string tachyon
and its variation to satisfy the equations of motion at $x^0=0$.
Our variational prescription imposes these conditions only at
the initial surface and allows off-shell configurations away from it.

In the interacting case, the symplectic form is obtained by replacing the BRST operator $Q$ with the kinetic operator $Q_{\Phi_{\mathrm{sol}}}$ of SFT expanded around a classical solution $\Phi_{\mathrm{sol}}$.
This relation may provide a useful guiding principle for constructing the corresponding boundary term in the interacting case.

We have exhibited two off-shell realizations of the additional term that restore cyclicity and reproduce the same tree-level two-point amplitudes.
The first is non-local in time, while the second is local in time and remains non-local in space.
For the open-string tachyon, the second realization gives a genuine boundary term and admits a variational prescription based on data at $x^0=0$.
The field and its variation satisfy the equations of motion at that surface, and $\delta T_-(0;\bm{x})=0$ fixes one component of the initial data.
These restrictions prepare initial data without imposing the equations of motion for all times.
This distinction between on-shell amplitudes and off-shell boundary prescriptions should be useful in studying string field theory with interfaces or boundaries.

\bigskip

\noindent
{\normalfont \bfseries \large Acknowledgments}

The work of H.~M. is supported by JSPS KAKENHI (Grant-in-Aid for Early-Career Scientists) grant number JP22K14038 and the Asahipen Hikari Foundation. 
The work of J.~T. is supported in part by JSPS KAKENHI Grant No. JP22H05115.
Y.~G. would like to
thank R. Adachi and S.~Chikazawa for helpful discussions.
\medskip

\appendix
\section{Details of the Calculations}

The component actions in this appendix use $P_\pm$ as defined in \eqref{P_pm}.

In this appendix, we collect the intermediate operators used in the derivation of the open- and closed-string actions in the presence of the operator $\Theta_0[X^0]$. The main purpose of this appendix is to make explicit how the action of $\Theta_0[X^0]$, the BRST operator $Q$, and the commutator $\big[Q,\Theta_0[X^0]\big]$ reorganize the string field into bulk and boundary contributions. In particular, the terms proportional to $\theta(x^0)$ reproduce the bulk part of the action, while the terms proportional to $\delta(x^0)$ arise from the discontinuity at the time slice $x^0=0$ and therefore contribute to the boundary part of the action.

\subsection{Open string}

We begin with the open-string sector. Our strategy is to first derive a set of auxiliary formulas for the action of $\Theta_0[X^0]$ and $\big[Q,\Theta_0[X^0]\big]$ on the component states appearing in the string field. We then substitute these identities into the symplectic form $\omega(\Phi,\Theta_0[X^0]Q\Phi)$ and into the boundary term involving $\{P_+,\big[Q,\Theta_0[X^0]\big]\}$. This separates the bulk contribution from the local surface terms and the time-nonlocal contribution involving $P_+$.

We first collect several formulas useful for deriving the open-string action. To obtain the action for the massless gauge field, one must eventually evaluate expressions of the form appearing in \eqref{[Q,theta]alpha_-1 c_1|k>}. As a first step, we examine the action of the step function. Concretely, this amounts to evaluating the relevant OPE.
\begin{align}
&\Theta_0[X^0]\alpha^\mu_{-1}\ket{0,k}\nonumber\\
&=\left(\oint_{\mathcal{C}}\frac{dz}{2\pi i}\frac{1}{z}\int_q\frac{i}{q}:e^{-iqX^0}(z,\bar{z}):+\frac{1}{2}
    \right)\frac{i}{\sqrt{2\alpha'}}:\partial X^\mu e^{ik\cdot X}(0,0):\ket{0}\notag\\[6pt]
    &=\frac{i}{\sqrt{2\alpha'}}\left[\oint_{\mathcal{C}}\frac{dz}{2\pi i}\frac{1}{z}
    \int_q \frac{i}{q}:e^{i\hat{q}\cdot X}(z,\bar{z}):\,:\!\partial X^\mu e^{ik\cdot X}(0,0)\!:\,+\,\frac{1}{2}:\!\partial X^\mu e^{ik\cdot X}(0,0)\!:
    \right]\ket{0}\notag\\[6pt]
    &=\frac{i}{\sqrt{2\alpha'}}\Big[
    \int_q\!\frac{i}{q}\left(\alpha'\hat{q}^\mu q:\!\partial X^0 e^{i(k+\hat{q})\cdot X}(0,0)\!:\,+\,:\!\partial X^\mu e^{i(k+\hat{q})\cdot X}(0,0)\!:
    \right)+ \frac{1}{2}:\!\partial X^\mu e^{ik\cdot X}(0,0)\!:
    \Big]\ket{0}\notag\\
    &=\int_q\!\frac{i}{q}\alpha^\mu_{-1}\ket{k+\hat{q}}+\frac{1}{2}\alpha^\mu_{-1}\ket{k}+\int_q i\alpha'\hat{q}^\mu \alpha^0_{-1}\ket{0,k+\hat{q}}\,\label{Theta alpha_-1 |k>}\,.
\end{align}
Here $\hat q$ is the vector whose zero-th component is $q$ and whose other components are all zero. Since the ghost mode $c_1$ commutes with the step function, we suppress it here. When the BRST operator $Q$ acts, however, one must keep in mind that it does not commute with $\alpha^\mu_{-1}$ and $c_1$:
\begin{align}
    \quad\Theta_0[X^0]Q\,\alpha^\mu_{-1} c_1\ket{0,k}
    &=
   \Theta_0[X^0]\left(
   \alpha^\mu_{-1}Q+\sum_{n=-\infty}^\infty c_{-n}\alpha^\mu_{n-1}
   \right)c_1\ket{0,k}\notag\\
   &=\Theta_0[X^0]\left(
   \alpha^\mu_{-1}\left(\alpha'k^2-1)\right)c_0c_1\ket{k}\right)
   +\Theta_0[X^0]\left(
   c_{-1}c_1\alpha^\mu_0+c_0c_1\alpha^\mu_{-1}
   \right)\ket{0,k}\notag\\
   &=\alpha'k^2c_0c_1\left(
   \frac{1}{2}\alpha^\mu_{-1}\ket{0,k}+\int_q\frac{i}{q}\alpha^\mu_{-1}\ket{0,k+\hat{q}}-\int_q i\alpha'q\, \eta^{\mu 0}\alpha^0_{-1}\ket{0,k+\hat{q}}   \right)\nonumber\\%\hat{q}^\mu=-q\eta^{\mu 0}
   &\quad+c_{-1}c_1\sqrt{2\alpha'}k^\mu \Theta_0[X^0]\ket{0,k}\,.
\end{align}
To obtain the second line, we evaluate the commutators of $Q$ in \eqref{BRST open} with $\alpha^\mu_{-1}$ and $c_1$, and then let the resulting expression act on $\ket{0,k}$. The third equality follows by ignoring the ghost sector and using \eqref{Theta alpha_-1 |k>}. Similarly, by computing the action of $Q\Theta_0[X^0]$ on $\alpha^\mu_{-1}c_1\ket{0,k}$, one obtains the action of
\begin{align}
    \left[Q,\Theta_0[X^0]\right]=Q\Theta_0[X^0]-\Theta_0[X^0]Q\notag
\end{align}
on $\alpha^\mu_{-1}c_1\ket{0,k}$:
\begin{align}
      &\left[Q,\Theta_0[X^0]\right] \alpha^\mu_{-1}c_1 \ket{0,k}\nonumber\\
      &=\int_q i\alpha'(-2k^0-q)c_0c_1\alpha^\mu_{-1}\ket{0,k+\hat{q}}\,+\,\int_q \hat{q}^\mu\Big[
      i\alpha'^2q(-2k^0-q)c_0c_1\alpha^0_{-1}\ket{0,k+\hat{q}}\notag\\[6pt]
      &\quad+\,\frac{i\sqrt{2\alpha'}}{q}c_{-1}c_1\ket{0,k+\hat{q}}\,+\,i\alpha'\sqrt{2\alpha'}(k^0+q)c_{-1}c_1\ket{0,k+\hat{q}}
      \Big]\label{[Q,theta]alpha_-1 c_1|k>}\,.
\end{align}
We next consider the action of $\left\{P_+,\big[Q,\Theta_0[X^0]\big]\right\}$. As discussed in Section~\ref{Sec:Definitions of some operators}, when $P_\pm$ acts on $\ket{k}$, it produces a step function in the energy. Because the anticommutator contains a term in which $\big[Q,\Theta_0[X^0]\big]$ acts first, one must take into account the shift of the energy from $k^0$ to $k^0+q$, as is evident from \eqref{[Q,theta]alpha_-1 c_1|k>}. With this understood, the result can be written as follows:
\begin{align}
      &\left\{P_+,\left[Q,\Theta_0[X^0]\right]
      \right\} \alpha^\mu_{-1} c_1 \ket{0,k}\nonumber\\
      &=\int_q\left(\theta(k^0)+\theta(k^0+q)\right)\times
      \Big(
       i\alpha'(-2k^0-q)c_0c_1\alpha^\mu_{-1}\ket{0,k+\hat{q}}\notag\\[6pt]
       &\quad+\, \hat{q}^\mu\Big[
      i\alpha'^2q(-2k^0-q)c_0c_1\alpha^0_{-1}\ket{0,k+\hat{q}}\notag\\[6pt]
      &\quad+\frac{i\sqrt{2\alpha'}}{q}c_{-1}c_1\ket{0,k+\hat{q}}\,+\,i\alpha'\sqrt{2\alpha'}(k^0+q)c_{-1}c_1\ket{0,k+\hat{q}}
      \Big]
      \Big)\,.
\end{align}
In the case of the auxiliary field $B$, the derivation of the action involves the following computation:
\begin{align}
    \hspace{0cm}\quad\Theta_0[X^0]Qc_0\ket{0,k}&=\Theta_0[X^0]\left(
    -c_0Q-\sum_{m=-\infty}^\infty m:\!c_{-m}c_m\!:
    \right)\ket{k}\notag\\[6pt]
    &=\Theta_0[X^0]\left(-c_0c_1\sqrt{2\alpha'}k_\mu \alpha^\mu_{-1}-2c_{-1}c_1
    \right)\ket{0,k}\notag\notag\\[6pt]
    &=-c_0c_1\sqrt{2\alpha'}k_\mu\left(
    \int_q\!\frac{i}{q}\alpha^\mu_{-1}\ket{0,k+\hat{q}}+\frac{1}{2}\alpha^\mu_{-1}\ket{0,k}+\int_q i\alpha'\hat{q}^\mu \alpha^0_{-1}\ket{0,k+\hat{q}}
    \right)\notag\\[6pt]
    &\quad-2c_{-1}c_1\left(
    \int_q\frac{i}{q}\ket{0,k+\hat{q}}+\frac{1}{2}\ket{0,k}
    \right)\,.
\end{align}
Similarly, one finds
\begin{align}    
\quad\left[Q,\Theta_0[X^0]\right]c_0\ket{0,k}=\int_q i\sqrt{2\alpha'}\left(1-\alpha'k^0q\right)\alpha^0_{-1}c_0c_1\ket{0,k+\hat{q}}\,.
\end{align}
In the symplectic pairing, these two structures lead respectively to the bulk term weighted by $\theta(x^0)$ and to the localized term weighted by $\delta(x^0)$. We now substitute the formulas above into the quadratic action and evaluate the resulting inner products explicitly. Using the results above, we can calculate the first term of the action as
\begin{align}
    &\omega \!\left(\,\Phi,\,\Theta_0[X^0] Q\,\Phi\right)\notag\\[4pt]
    &=\int_{k_1}\!\int_{k_2}\left(\bra{0,k_1}c_{-1}\alpha^\nu_1A_\nu(k_1)\,+\,i\sqrt{\frac{\alpha'}{2}}\bra{0,k_1}c_0B(k_1)\right)\,\Theta_0[X^0]Q\notag\\[6pt]
    &\quad\times\,\left(A_\mu(k_2)c_{1}\alpha^\mu_{-1}\ket{0,k_2}\,-\,i\sqrt{\frac{\alpha'}{2}}B(k_2)c_0\ket{0,k_2}\right)\notag\\
    &=\int_{k_1}\!\int_{k_2}A_\mu(k_2)A_\nu(k_1)\bra{0,k_1}\left\{\alpha'k_2^2\left(\frac{1}{2}\eta^{\mu\nu}\ket{0,k_2}\,+\,\int_q\!\frac{i}{q}\eta^{\mu\nu}\ket{0,k_2+\hat{q}}\,-\,\int_q i\alpha'q \eta^{\mu0}\eta^{\nu0}\ket{0,k_2+\hat{q}}\right)\right\}\notag\\[6pt]
    &\quad+{\alpha'}\!\int_{k_1}\!\int_{k_2}B(k_1)B(k_2)\bra{0,k_1}\Big(\int_q\!\frac{i}{q}\ket{0,k_2+\hat{q}}+\frac{1}{2}\ket{0,k_2}\Big)\notag\\[6pt]
    &\quad+i\alpha'\int_{k_1}\!\int_{k_2}\big(A_\mu(k_1)B(k_2)-A_\mu(k_2)B(k_1)\,\big)k_2^\mu\bra{0,k_1}\left(\int_q \!\frac{i}{q}\ket{0,k_2+\hat{q}}\,+\,\frac{1}{2}\ket{0,k_2}\right)\notag\\[6pt]
    &\quad+i\sqrt{\frac{\alpha'}{2}}\int_{k_1}\!\int_{k_2}A_\nu(k_1)B(k_2)\bra{0,k_1}\int_q\!(-i\alpha')q\eta^{\mu0}\eta^{\nu0}\sqrt{2\alpha'}(k_2)_\mu\ket{0,k_2+\hat{q}}\notag\\
    &= \int_{k_1}\!\int_{k_2}\!\Big\{A_\mu(k_1)A^\mu(k_2)\alpha'k_2^2+\alpha'B(k_1)B(k_2)\notag\\[6pt]
    &\quad +i\alpha'k_2^\mu\left(A_\mu(k_1)B(k_2)-A_\mu(k_2)B(k_1)\,\right)\Big\}\!\int d^dx \,\theta(x^0)e^{i(k_1+k_2)\cdot x}\notag\\[6pt]
    &\quad-i\alpha'\!\int_{k_1}\!\int_{k_2}\left(\alpha'k_2^2 A^0(k_1)A^0(k_2)+i\alpha'k_2^0 A^0(k_1)B(k_2)\right)(-k_1^0-k_2^0)\,\!\int d^dx \,\delta(x^0) e^{i(k_1+k_2)\cdot x}\,.
    \label{massless open bulk}
\end{align}
Equation \eqref{massless open bulk} shows that the contribution from $\omega(\Phi,\Theta_0[X^0]Q\Phi)$ naturally splits into two parts. The terms in the first and second lines on the right-hand side of the last equation constitute the bulk contribution. They are supported in the region $x^0>0$ and reproduce the expected quadratic kinetic structure for the component fields $A_\mu$ and $B$. The third line is proportional to $\delta(x^0)$ and is therefore localized on the time boundary. 
There is another surface contribution computed as
\begin{align}
    &\omega(\,\Phi,\tfrac{1}{2}\left\{P_+,\left[Q,\Theta_0[X^0]\right]\right\}\,\Phi)\notag\\[4pt]
    &=\int_{k_1}\!\int_{k_2}\Big\{A_\mu(k_1)A^\mu(k_2)i\alpha'(k_1^0-k_2^0)-i\alpha'^2A^0(k_1)A^0(k_2)(k_1^0-k_2^0)(k_1^0+k_2^0)^2\notag\\[6pt]
    &\qquad\qquad\quad+\alpha'A^0(k_1)B(k_2)\big(1+\alpha'k_2^0(k_1^0+k_2^0)\big)\notag\\[6pt]
    &\qquad\qquad\quad-\alpha'A^0(k_2)B(k_1)\big(1+\alpha'k_1^0(k_1^0+k_2^0)\big)\Big\}\notag\\[6pt]
    &\hspace{5.8cm}\times\frac{\theta(-k_1^0)+\theta(k_2^0)}{2}\int d^dx \,\delta(x^0)e^{i(k_1+k_2)\cdot x}\,.\label{massless open boundary}
\end{align}
The term in \eqref{massless open boundary} is the additional boundary contribution coming from the anticommutator $\{P_+,\big[Q,\Theta_0[X^0]\big]\}$. For the choice \eqref{P_pm}, the energy step functions make this contribution non-local in time despite the factor $\delta(x^0)$. Together with the $\delta(x^0)$ part already present in \eqref{massless open bulk}, it determines the full boundary part in the open-string action. Writing the result in this way makes transparent which part of the quadratic action is due to the bulk dynamics and which part is induced by the spacelike interface at $x^0=0$.

\subsection{Closed string}

We next turn to the closed-string sector. As in the open-string case, the calculation is organized in two steps. First, we collect the action of $Q$, $\Theta_0[X^0]$, and $\big[Q,\Theta_0[X^0]\big]$ on the basis states that appear in the decomposition of the string field. Second, we insert these identities into the symplectic pairing and separate the result into bulk and boundary pieces. Because the closed-string field contains several component fields, including $e_{\mu\nu}$, $e_\mu$, $\bar e_\mu$, $e$, and $\bar e$, it is useful to keep track of how each operator contributes to the mixing among these components. 
The computation is almost identical to that in the open-string case, but it is somewhat more involved due to the presence of antiholomorphic modes.

We begin by examining how the BRST operator $Q$ in \eqref{BRST closed} acts on each of the closed-string states:
\begin{align}
    &\quad Q\alpha^\mu_{-1}\tilde{\alpha}^\nu_{-1}c_1\tilde{c}_1\ket{0,k}=\frac{\alpha'k^2}{4}2c_0^+c_1\tilde{c}_1\alpha^\mu_{-1}\tilde{\alpha}^\nu_{-1}\ket{0,k}\,+\,c_{-1}c_1\tilde{c}_1\alpha^\mu_0\tilde{\alpha}^\nu_{-1}\ket{0,k}\,+\,\tilde{c}_{-1}c_1\tilde{c}_1\alpha^\mu_{-1}\tilde{\alpha}^\nu_{0}\ket{0,k}\,,
\end{align}
\vspace{-0.8cm}
\begin{align}
    &\quad Qc_0^+c_1\alpha^\mu_{-1}\ket{0,k}=\sqrt{\frac{\alpha'}{2}}k_\nu\alpha^\mu_{-1}\tilde{\alpha}^\nu_{-1}c_0^+c_1\tilde{c}_1\ket{0,k}\,+\,\sqrt{\frac{\alpha'}{2}}k^\mu c_0^+c_1{c}_{-1}\ket{0,k}\,+\,\tilde{c}_{-1}c_1\tilde{c}_1\alpha^\mu_{-1}\ket{0,k}\,,
\end{align}
\vspace{-0.8cm}
\begin{align}
    &\quad Qc_{-1}c_{1}\ket{0,k}=\frac{\alpha'k^2}{4}2c_0^+c_{-1}{c}_1\ket{0,k}\,+\,\sqrt{\frac{\alpha'}{2}}k_\mu c_{-1}c_1\tilde{c}_1\tilde{\alpha}^\mu_{-1}\ket{0,k}\,.
\end{align}
Next, we consider the action of the step function $\Theta_0[X^0]$. As in the open-string case, it acts trivially on the ghost sector. Taking into account the presence of anti-holomorphic modes, one finds
\begin{align}
\quad \Theta_0[X^0]\alpha^\mu_{-1}\tilde{\alpha}^\nu_{-1}c_1\tilde{c}_1\ket{0,k}
    =&\,c_1\tilde{c}_1\Big(\frac{1}{2}\alpha^\mu_{-1}\tilde{\alpha}^\nu_{-1}\ket{k}\,+\,\int_q\!\frac{i}{q}\alpha^\mu_{-1}\tilde{\alpha}^\nu_{-1}\ket{0,k+\hat{q}}\,\notag\\[6pt]
    &\hspace{3cm}+\,\int_q\!\frac{i\alpha'}{2}(\hat{q}^\mu \alpha^0_{-1}\tilde{\alpha}^\nu_{-1}\,+\,\hat{q}^\nu\alpha^\mu_{-1}\tilde{\alpha}^0_{-1})\ket{0,k+\hat{q}}\notag\\
    &\hspace{3cm}+\int_q\frac{i\alpha'^2}{4}\,
 q\,\hat q^\mu\hat q^\nu
 \alpha^0_{-1}\tilde\alpha^0_{-1}
 \ket{0,k+\hat q}\Big)\,,
\end{align}
\vspace{-5mm}
\begin{align}
    \quad\Theta_0[X^0]\alpha^\mu_{-1}c_1\ket{0,k}=c_1\left(\frac{1}{2}\alpha^\mu_{-1}\ket{0,k}\,+\,\int_q\!\frac{i}{q}\alpha^\mu_{-1}\ket{0,k+\hat{q}}\,+\,\frac{1}{2}\int_q\!i\alpha'\hat{q}^\mu\alpha^0_{-1}\ket{0,k+\hat{q}}\right)\,,
\end{align}
\vspace{-5mm}
\begin{align}
    \quad\Theta_0[X^0]\ket{0,k}=\frac{1}{2}\ket{0,k}+\int_q \!\frac{i}{q}\ket{0,k+\hat{q}}\,.
\end{align}
Combining these results, one can derive the formulas required for the kinetic term in the action as follows:
\vspace{-0.4cm}
\begin{align}
&\left[Q,\Theta_0[X^0]\right]\alpha^\mu_{-1}\tilde{\alpha}^\nu_{-1}c_1\tilde{c}_1\ket{0,k}\nonumber\\
    &=\int_q\!i\alpha'\frac{-2k^0-q}{4}\Big[\alpha^\mu_{-1}\tilde{\alpha}^\nu_{-1}-\frac{\alpha'q^2}{2}\big(\eta^{\mu0}\alpha^0_{-1}\tilde{\alpha}^\nu_{-1}+\eta^{\nu0}\alpha^\mu_{-1}\tilde{\alpha}_{-1}^0\big)\notag\\
    &\hspace{6.5cm}+\frac{\alpha'^2q^4}{4}\eta^{\mu0}\eta^{\nu0}\alpha^0_{-1}\tilde{\alpha}^0_{-1}
    \Big]
    2c_0^+c_1\tilde{c}_1\ket{0,k+\hat{q}}\notag\\[6pt]
    &\quad-i\int_q\sqrt{\frac{\alpha'}{2}}\,
 \eta^{\mu0}
 \left(1+\frac{\alpha'q(k^0+q)}{2}\right)
 \left(
 \tilde\alpha^\nu_{-1}
 -\frac{\alpha'q^2}{2}\eta^{\nu0}\tilde\alpha^0_{-1}
 \right)
 c_{-1}c_1\tilde c_1\ket{0,k+\hat q}
 \notag\\
&\quad-i\int_q\sqrt{\frac{\alpha'}{2}}\,
 \eta^{\nu0}
 \left(1+\frac{\alpha'q(k^0+q)}{2}\right)
 \left(
 \alpha^\mu_{-1}
 -\frac{\alpha'q^2}{2}\eta^{\mu0}\alpha^0_{-1}
 \right)
 \tilde c_{-1}c_1\tilde c_1\ket{0,k+\hat q}\,,
\end{align}
\vspace{-0.8cm}
\begin{align}
    &\hspace{-1cm}\left[Q,\Theta_0[X^0]\right]c_0^+c_1\alpha^\mu_{-1}\ket{0,k}\nonumber\\
    &=-i\int_q\sqrt{\frac{\alpha'}{2}}\,
 \left(1-\frac{\alpha'qk^0}{2}\right)
 \left(
 \alpha^\mu_{-1}
 -\frac{\alpha'q^2}{2}\eta^{\mu0}\alpha^0_{-1}
 \right)
 \tilde\alpha^0_{-1}
 c_0^+c_1\tilde c_1\ket{0,k+\hat q}\notag\\[6pt]
    &\hspace{3.1cm}-i\!\int_q\!\sqrt{\frac{\alpha'}{2}}\eta^{\mu0}\left(1+\frac{\alpha'q}{2}(k^0+q)\right)c_0^+c_1c_{-1}\ket{0,k+\hat{q}}\,,
\end{align}
\vspace{-0.5cm}
\begin{align}
    &\left[Q,\Theta_0[X^0]\right]c_{-1}c_{1}\ket{0,k}\nonumber\\
    &=\int_q\!i\alpha'\frac{-2k^0-q}{4}2c_0^+c_{-1}c_1\ket{k+\hat{q}}\,-\,\int_q\!i\sqrt{\frac{\alpha'}{2}}\left(1-\frac{\alpha'qk^0}{2}\right)c_{-1}c_1\tilde{c}_1\tilde{\alpha}^0_{-1}\ket{0,k+\hat{q}}\,.
\end{align}
The formulas listed above play the same role as in the open-string sector. The action of $\Theta_0[X^0]$ generates the bulk contribution together with a boundary-supported term, while the commutator $\big[Q,\Theta_0[X^0]\big]$ produces the additional surface contribution that must be added separately. The following computation makes this decomposition explicit at the level of the component fields. We perform the calculation of the first term of the action:
\begin{align}
    \hspace{1cm}&\omega(\Phi,\Theta_0[X^0]Q \,\Phi)\notag\\
    &=\int_{k_1}\!\int_{k_2}\Bigg(
     \bra{0,k_1}\tilde{c}_{-1}c_{-1}\tilde{\alpha}^\lambda_1\alpha^\rho_1 \frac{-1}{2}e_{\rho\lambda}(k_1)+\bra{0,k_1}\left(e(k_1)c_1c_{-1}+\bar{e}(k_1)\tilde{c}_1\tilde{c}_{-1}\right)\notag\\
     &\hspace{3cm}-i\sqrt{\frac{\alpha'}{2}}\bra{0,k_1}\left(c_{-1}c_0^+\alpha^\nu_1e_\nu(k_1)+\tilde{c}_{-1}c_0^+\tilde{\alpha}^\nu_{1}\bar{e}_\nu(k_1)\right)
    \Bigg)\,c_0^-\Theta_0[X^0]Q\notag\\
    &\hspace{0.4cm}\times\Bigg\{-\frac{1}{2}
    e_{\mu\nu}(k_2)c_1\tilde{c}_1\alpha^\mu_{-1}\tilde{\alpha}^\nu_{-1}\ket{0,k_2}+\left(e(k_2)c_1c_{-1}+\bar{e}(k_2)\tilde{c}_1\tilde{c}_{-1}\right)\ket{0,k_2}\notag\\
    &\hspace{5cm}+i\sqrt{\frac{\alpha'}{2}}c_0^+\left(
    e_\mu(k_2)c_1\alpha^\mu_{-1}+\bar{e}_\mu(k_2) \tilde{c}_1\tilde{\alpha}^\mu_{-1}\right)\ket{0,k_2}
    \Bigg\}\notag\\
&=\int_{k_1}\!\int_{k_2}\!\int\!d^dx\,\theta(x^0)e^{i(k_1+k_2)\cdot x}\,\frac{\alpha'}{4}\Bigg[\frac{
    k_2^{\,2}}{4}e_{\mu\nu}(k_1)e^{\mu\nu}(k_2)\,-\frac{i}{2}\,e_{\mu\nu}(k_1)e^\mu(k_2)k_2^\nu\,+\frac{i}{2}\,e_{\mu\nu}(k_2)e^\mu(k_1)k_2^\nu\notag\\[6pt]
    &\hspace{4cm}+\frac{i}{2}e_{\mu\nu}(k_1)\bar{e}^\nu(k_2)k_2^\mu\,-\frac{i}{2}\,e_{\mu\nu}(k_2)\bar{e}^\nu(k_1)k_2^\mu\,+\,k_2^2\big(\,e(k_1)\bar{e}(k_2)+e(k_2)\bar{e}(k_1)\,\big)\notag\\[6pt]
    &\hspace{4cm}+ie(k_1)\bar{e}_\mu(k_2)k_2^\mu\,-\,ie(k_2)\bar{e}_\mu(k_1)k_2^\mu\,+\,i\bar{e}(k_1)e_\mu(k_2)k_2^\mu\,-\,i\bar{e}(k_2)e_\mu(k_1)k_2^\mu\notag\\[6pt]
    &\hspace{11cm}+e_\mu(k_1)e^\mu(k_2)\,+\,\bar{e}^\mu(k_1)\bar{e}_\mu(k_2)\,\Bigg]\notag
\end{align}
\vspace{-0.8cm}
\begin{align}
&\hspace{1cm}+\int_{k_1}\!\int_{k_2}\!\int\!d^dx\,\delta(x^0) e^{i(k_1+k_2)\cdot x}\,\frac{i\alpha'^2}{8}(k_1^0+k_2^0)\notag\\[6pt]
    &\hspace{3cm}\times\Bigg[\,
    \eta^{\mu\nu\rho\lambda}\frac{k_2^{\,2}}{4} e_{\rho\lambda}(k_1)e_{\mu\nu}(k_2)\,-\frac{i}{2}\,\eta^{\mu\nu\rho\lambda}e_{\rho\lambda}(k_1)e_\mu(k_2){k_2}_\nu+\frac{i}{2}e^{0\nu}(k_2)e^0(k_1){k_2}_\nu\notag\\
    &\hspace{4.4cm}+\frac{i}{2}e_{\rho\lambda}(k_1)\bar{e}_\nu(k_2){k_2}_\mu\eta^{\mu\nu\rho\lambda}\,-\,\frac{i}{2}{k_2}_\mu e^{\mu0}(k_2)\bar{e}^0(k_1)\notag\\
    &\hspace{4.4cm}-i\bar{e}^0(k_1)e(k_2)k_2^0-ie^0(k_1)\bar{e}(k_2)k_2^0
    \,+\,{e}^0(k_1)e^0(k_2)\,+\,\bar{e}^0(k_1)\bar{e}^0(k_2)\,
    \Bigg]\notag\\
    &\quad-\frac{i\alpha'^3}{16}
 \int_{k_1}\int_{k_2}\int d^dx\,
 \delta(x^0)e^{i(k_1+k_2)\cdot x}
 (k_1^0+k_2^0)^3 e^{00}(k_1)\left[
 \frac{k_2^2}{4}e^{00}(k_2)
 -\frac{i}{2}k_2^0e^0(k_2)
 +\frac{i}{2}k_2^0\bar e^0(k_2)
 \right]\,.\label{closed massless bulk}
\end{align}
The result in \eqref{closed massless bulk} already exhibits the basic structure of the closed-string action in the present background. The part proportional to $\theta(x^0)$ gives the bulk quadratic form for the component fields, while the part proportional to $\delta(x^0)$ is localized on $x^0=0$. The tensor $\eta^{\mu\nu\rho\lambda}$ introduced below is a convenient shorthand
$\eta^{\mu\nu\rho\lambda} := \eta^{\mu 0} \eta^{\rho 0} \eta^{\nu \lambda} \,+\, \eta^{\nu 0} \eta^{\lambda 0} \eta^{\mu \rho}$.
Furthermore, the part corresponding to the surface term of the action can be written as follows:
\begin{align}
    &\omega(\Phi,\frac{1}{2}\left\{P_+,\left[Q,\Theta_0[X^0]\right]\right\}\Phi)\notag\\[4pt]
    &=\int_{k_1}\!\int_{k_2}\Bigg(
     \bra{0,k_1}\tilde{c}_{-1}c_{-1}\tilde{\alpha}^\lambda_1\alpha^\rho_1 \frac{-1}{2}e_{\rho\lambda}(k_1)+\bra{0,k_1}\left(e(k_1)c_1c_{-1}+\bar{e}(k_1)\tilde{c}_1\tilde{c}_{-1}\right)\notag\\[1pt]
     &\qquad\qquad\qquad-i\sqrt{\frac{\alpha'}{2}}\bra{0,k_1}\left(c_{-1}c_0^+\alpha^\nu_1e_\nu(k_1)+\tilde{c}_{-1}c_0^+\tilde{\alpha}^\nu_{1}\bar{e}_\nu(k_1)\right)
    \Bigg)\,c_0^-\frac{1}{2}\left\{P_+,\left[Q,\Theta_0[X^0]\right]\right\}\notag\\[4pt]
    &\hspace{0.7cm}\times\Big(\frac{-1}{2}
    e_{\mu\nu}(k_2)c_1\tilde{c}_1\alpha^\mu_{-1}\tilde{\alpha}^\nu_{-1}\ket{0,k_2}+\left(e(k_2)c_1c_{-1}+\bar{e}(k_2)\tilde{c}_1\tilde{c}_{-1}\right)\ket{k_2}\notag\\
    &\hspace{6cm}+i\sqrt{\frac{\alpha'}{2}}c_0^+\left(
    e_\mu(k_2)c_1\alpha^\mu_{-1}+\bar{e}_\mu(k_2) \tilde{c}_1\tilde{\alpha}^\mu_{-1}\right)\ket{0,k_2}
    \Big)\notag
\end{align}
\begin{align}
    &=\int_{k_1}\!\int_{k_2}\frac{\theta(k_2^0)+\theta(-k_1^0)}{2}\,\int\!d^dx \,\delta(x^0)e^{i(k_1+k_2)\cdot x}\,\frac{\alpha'}{4}\notag\\[6pt]
    &\hspace{1.1cm}\times\Bigg[
    i\frac{k_1^0-k_2^0}{4}\left(\eta^{\mu\rho}\eta^{\nu\lambda}\,-\,\frac{\alpha'(k_1^0+k_2^0)^2}{2}\eta^{\mu\nu\rho\lambda}
    +\frac{\alpha'^2(k_1^{\,0}+k^{\,0}_2)^4}{4}\eta^{\mu0}\eta^{\nu0}\eta^{\rho0}\eta^{\lambda0}
    \right)e_{\rho\lambda}(k_1)e_{\mu\nu}(k_2)\notag\\[6pt]
    %%%%%%%%%%%%%%%%%%%%%%%%%
    &\hspace{3cm}-\frac12
 \left(1+\frac{\alpha'k_2^0(k_1^0+k_2^0)}{2}\right)
 \left(
 \eta^{\mu\rho}\eta^{\lambda0}
 -\frac{\alpha'(k_1^0+k_2^0)^2}{2}
  \eta^{\mu0}\eta^{\rho0}\eta^{\lambda0}
 \right)
 e_{\rho\lambda}(k_1)e_\mu(k_2)
 \notag\\
&\hspace{3cm}+\frac12
 \left(1+\frac{\alpha'k_2^0(k_1^0+k_2^0)}{2}\right)
 \left(
 \eta^{\mu\lambda}\eta^{\rho0}
 -\frac{\alpha'(k_1^0+k_2^0)^2}{2}
  \eta^{\mu0}\eta^{\lambda0}\eta^{\rho0}
 \right)
 e_{\rho\lambda}(k_1)\bar e_\mu(k_2)
 \notag\\
&\hspace{3cm}+\frac12
 \left(1+\frac{\alpha'k_1^0(k_1^0+k_2^0)}{2}\right)
 \left(
 \eta^{\mu\rho}\eta^{\nu0}
 -\frac{\alpha'(k_1^0+k_2^0)^2}{2}
  \eta^{\mu0}\eta^{\rho0}\eta^{\nu0}
 \right)
 e_{\mu\nu}(k_2)e_\rho(k_1)
 \notag\\
&\hspace{3cm}-\frac12
 \left(1+\frac{\alpha'k_1^0(k_1^0+k_2^0)}{2}\right)
 \left(
 \eta^{\nu\rho}\eta^{\mu0}
 -\frac{\alpha'(k_1^0+k_2^0)^2}{2}
  \eta^{\nu0}\eta^{\rho0}\eta^{\mu0}
 \right)
 e_{\mu\nu}(k_2)\bar e_\rho(k_1)
 \notag\\
 %%%%%%%%%%%%%%%%%%%%%%%%%%%%%%%
    &\hspace{3cm}+i(k_1^0-k_2^0)\big(\,e(k_1)\bar{e}(k_2)\,+\,e(k_2)\bar{e}(k_1)\,\big)\notag\\[6pt]
    &\hspace{6cm}+\Big(1+\frac{\alpha'k_1^0(k_1^0+k_2^0)}{2}\Big)\eta^{\mu0}\big(\,e(k_1)\bar{e}_\mu(k_2)+\bar{e}(k_1)e_\mu(k_2)\,\big)\notag\\[6pt]
    &\hspace{6cm}-\Big(1+\frac{\alpha'k_2^0(k_1^{\,0}+k_2^{\,0})}{2}\Big)\eta^{\nu0}\big(\,e(k_2)\bar{e}_\nu(k_1)+\bar{e}(k_2)e_\nu(k_1)\,\big)\,
    \Bigg]\,.\label{Closed action 2nd term}
\end{align}
Equation \eqref{Closed action 2nd term} gives the remaining surface contribution. As in the open-string case, the factor $\delta(x^0)$ does not localize the term containing the energy step functions in time. When combined with the $\delta(x^0)$ term appearing in \eqref{closed massless bulk}, it yields the complete boundary part of the quadratic closed-string action. Presenting the calculation in this form clarifies that the boundary couplings are not introduced by hand, but follow systematically from the commutator of the BRST operator with operator $\Theta_0[X^0]$.

% Appendix B has been moved to Section 5; its original deletions are marked there.


\begin{thebibliography}{999}

%\cite{Firat:2024kxq}
\bibitem{Firat:2024kxq}
A.~H.~F{\i}rat and R.~A.~Mamade,
``Boundary terms in string field theory,''
JHEP \textbf{02} (2025), 058
doi:10.1007/JHEP02(2025)058
[arXiv:2411.16673 [hep-th]].
%13 citations counted in INSPIRE as of 07 Feb 2026

%\cite{Stettinger:2024uus}
\bibitem{Stettinger:2024uus}
G.~Stettinger,
``A boundary term for open string field theory,''
JHEP \textbf{05} (2025), 226
doi:10.1007/JHEP05(2025)226
[arXiv:2411.15123 [hep-th]].
%11 citations counted in INSPIRE as of 07 Feb 2026

%\cite{Maccaferri:2025orz}
\bibitem{Maccaferri:2025orz}
C.~Maccaferri, R.~Poletti, A.~Ruffino and J.~Vo{\v{s}}mera,
``Boundary modes in string field theory,''
JHEP \textbf{06} (2025), 108
doi:10.1007/JHEP06(2025)108
[arXiv:2502.19373 [hep-th]].
%10 citations counted in INSPIRE as of 20 Dec 2025

%\cite{Maccaferri:2025onc}
\bibitem{Maccaferri:2025onc}
C.~Maccaferri, A.~Ruffino and J.~Vo{\v{s}}mera,
``Gauge-invariant action for free string field theory with boundary,''
[arXiv:2506.05969 [hep-th]].
%5 citations counted in INSPIRE as of 31 Dec 2025

%\cite{Alfonsi:2024utl}
\bibitem{Alfonsi:2024utl}
L.~Alfonsi, L.~Borsten, H.~Kim, M.~Wolf and C.~A.~S.~Young,
``Full S-matrices and witten diagrams with relative L$_\infty$-algebras,''
JHEP \textbf{07} (2025), 267
doi:10.1007/JHEP07(2025)267
[arXiv:2412.16106 [hep-th]].
%4 citations counted in INSPIRE as of 20 Dec 2025

%\cite{Cattaneo:2012qu}
\bibitem{Cattaneo:2012qu}
A.~S.~Cattaneo, P.~Mnev and N.~Reshetikhin,
``Classical BV theories on manifolds with boundary,''
Commun. Math. Phys. \textbf{332} (2014), 535-603
doi:10.1007/s00220-014-2145-3
[arXiv:1201.0290 [math-ph]].
%129 citations counted in INSPIRE as of 29 Apr 2026

%\cite{Erbin:2019uiz}
\bibitem{Erbin:2019uiz}
H.~Erbin, J.~Maldacena and D.~Skliros,
``Two-Point String Amplitudes,''
JHEP \textbf{07}, 139 (2019)
doi:10.1007/JHEP07(2019)139
[arXiv:1906.06051 [hep-th]].
%51 citations counted in INSPIRE as of 31 Dec 2025

%\cite{Seki:2019ycz}
\bibitem{Seki:2019ycz}
S.~Seki and T.~Takahashi,
``Two-point String Amplitudes Revisited by Operator Formalism,''
Phys. Lett. B \textbf{800}, 135078 (2020)
doi:10.1016/j.physletb.2019.135078
[arXiv:1909.03672 [hep-th]].
%13 citations counted in INSPIRE as of 31 Dec 2025

%\cite{Seki:2021ivm}
\bibitem{Seki:2021ivm}
S.~Seki and T.~Takahashi,
``Reduction of open string amplitudes by mostly BRST exact operators,''
Phys. Lett. B \textbf{822}, 136664 (2021)
doi:10.1016/j.physletb.2021.136664
[arXiv:2108.05628 [hep-th]].
%4 citations counted in INSPIRE as of 31 Dec 2025

%\cite{Kishimoto:2024aig}
\bibitem{Kishimoto:2024aig}
I.~Kishimoto, S.~Seki and T.~Takahashi,
``Two-point closed string amplitudes in the BRST formalism,''
Phys. Lett. B \textbf{853} (2024), 138657
doi:10.1016/j.physletb.2024.138657
[arXiv:2402.07464 [hep-th]].
%2 citations counted in INSPIRE as of 20 Jan 2026


%\cite{Bernardes:2025uzg}
\bibitem{Bernardes:2025uzg}
V.~Bernardes, T.~Erler and A.~H.~F{\i}rat,
``Covariant phase space and L$_\infty$ algebras,''
JHEP \textbf{09} (2025), 057
doi:10.1007/JHEP09(2025)057
[arXiv:2506.20706 [hep-th]].
%4 citations counted in INSPIRE as of 20 Dec 2025

%\cite{Bernardes:2025zzu}
\bibitem{Bernardes:2025zzu}
V.~Bernardes, T.~Erler and A.~H.~F{\i}rat,
``Symplectic structure in open string field theory. Part I. Rolling tachyons,''
JHEP \textbf{02} (2026), 063
doi:10.1007/JHEP02(2026)063
[arXiv:2511.03777 [hep-th]].
%9 citations counted in INSPIRE as of 30 Apr 2026

%\cite{Bernardes:2025zkj}
\bibitem{Bernardes:2025zkj}
V.~Bernardes, T.~Erler and A.~H.~F{\i}rat,
``Symplectic structure in open string field theory. Part II. Sliding lump,''
JHEP \textbf{02} (2026), 064
doi:10.1007/JHEP02(2026)064
[arXiv:2511.15781 [hep-th]].
%6 citations counted in INSPIRE as of 30 Apr 2026

%\cite{Bernardes:2026egp}
\bibitem{Bernardes:2026egp}
V.~Bernardes, T.~Erler and A.~H.~F{\i}rat,
``Symplectic structure in open string field theory III: Electric field,''
[arXiv:2604.01273 [hep-th]].
%4 citations counted in INSPIRE as of 30 Apr 2026









\bibitem{Ali:2026BEF}

M.~Ali and G.~Stettinger,
``The BEF symplectic form: a Lagrangian perspective,''
JHEP \textbf{07} (2026), 222
doi:10.1007/JHEP07(2026)222
[arXiv:2604.07334 [hep-th]].


\end{thebibliography}
\end{document}